\documentclass[aps,pra,twocolumn,footinbib,superscriptaddress]{revtex4-2} 
	
	\usepackage{mathtools,amsmath}
	\usepackage{graphicx} 
	\usepackage[breaklinks=true,colorlinks,citecolor=blue,linkcolor=blue,urlcolor=blue]{hyperref}
	\usepackage{physics}
	\usepackage{siunitx}
	\usepackage{bbold}
	\usepackage{soul}
	\usepackage{bm}
	\usepackage[normalem]{ulem}
	\usepackage{enumerate}  
	\usepackage{float}
	\usepackage{amssymb}

	\usepackage{xcolor}

\begin{document}
\title{Pareto-optimal cycles for power, efficiency and fluctuations of quantum heat engines using reinforcement learning}

\author{Paolo A. Erdman}
\affiliation{Freie Universit{\" a}t Berlin, Department of Mathematics and Computer Science, Arnimallee 6, 14195 Berlin, Germany}

\author{Alberto Rolandi}
\affiliation{Department of Applied Physics, University of Geneva, 1211 Geneva, Switzerland}

\author{Paolo Abiuso}
\affiliation{ICFO – Institut de Ci\`encies Foto\`niques, The Barcelona Institute of Science and Technology, 08860 Castelldefels (Barcelona), Spain}

\author{Mart\' i Perarnau-Llobet}
\email{marti.perarnaullobet@unige.ch}
\affiliation{Department of Applied Physics, University of Geneva, 1211 Geneva, Switzerland}

\author{Frank No{\'e}}
\email{frank.noe@fu-berlin.de}
\affiliation{Microsoft Research AI4Science, Karl-Liebknecht Str. 32, 10178 Berlin, Germany}
\affiliation{Freie Universit{\" a}t Berlin, Department of Mathematics and Computer Science, Arnimallee 6, 14195 Berlin, Germany}
\affiliation{Freie Universit{\" a}t Berlin, Department of Physics, Arnimallee 6, 14195 Berlin, Germany}
\affiliation{Rice University, Department of Chemistry, Houston, TX 77005, USA}

\begin{abstract}
The full optimization of a quantum heat engine requires operating at high power, high efficiency, and high stability (i.e. low power fluctuations). 
However, these three objectives cannot be simultaneously optimized - as indicated by the so-called thermodynamic uncertainty relations - and a systematic approach to finding optimal balances between them including power fluctuations has, as yet, been elusive. Here we propose such a general framework to identify Pareto-optimal cycles for driven quantum heat engines that trade-off power, efficiency, and fluctuations. We then employ reinforcement learning to identify the Pareto front of a quantum dot based engine and find abrupt changes in the form of optimal cycles when switching between optimizing two and three objectives. We further derive analytical results in the fast and slow-driving regimes that accurately describe different regions of the Pareto front.
\end{abstract}
\maketitle

\paragraph*{Introduction.}
Stochastic heat engines are devices that convert between heat and work at the nanoscale~\cite{giazotto2006,pekola2015,vinjanampathy2016}. Steady-state heat engines (SSHE) perform work against external thermodynamic forces (e.g. a chemical potential difference) after reaching a non-equilibrium steady state~\cite{benenti2017}, while periodically driven heat engines (PDHE) perform work against external driving fields through time-dependent cycles~\cite{binder2019}. 
The performance of 
heat engines is usually characterized by the output power and efficiency, and their optimization has been thoroughly addressed in literature \cite{geva1992,feldmann2000,schmiedl2008,esposito2010_prl,esposito2010_pre,allahverdyan2013,juergens2013,Holubec2015,Holubec2016,karimi2016,campisi2016,yamamoto2016,Kosloff2017,patel2017,cavina2018,abiuso2018,Ma2018,erdman2019_njp,menczel2019_prb,chen2019,abiuso2020_prl,cavina2021,hernandez2021,alonso2022geometric,Ye2022}.
However, in contrast to their macroscopic counterpart, the performance of quantum and microscopic engines is strongly influenced by power fluctuations.
While early works have started optimizing power fluctuation \cite{Holubec2014,Miller2020a,Denzler2021,Saryal2021b,Mehboudi2021}, a framework to fully optimize the performance of microscopic heat engines that accounts for power fluctuations is currently lacking; this letter fills this void.

An ideal engine operates at high power, high efficiency, and low power fluctuations; however, such quantities usually cannot be optimized simultaneously, but one must seek trade-offs. In SSHEs, a rigorous manifestation of this trade-off is given by the thermodynamic uncertainty relations~\cite{barato2015,gingrich2016,pietzonka2018,guarnieri2019,timpanaro2019,falasco2020,friedman2020,horowitz2020,saryal2021}. For ``classical'' stochastic SSHE (i.e. in the absence of quantum coherence) operating between two Markovian reservoirs at inverse
temperatures $\beta_\text{C}$ (cold) and  $\beta_\text{H}$ (hot), they read~\cite{pietzonka2018}:
\begin{equation}
    \xi \equiv \frac{2}{\beta_\text{C}} \frac{\ev*{P}}{\ev*{\Delta P} }\frac{\eta}{\eta_\text{c}-\eta} \leq 1,
    \label{eq:tur_ineq}
\end{equation}
where $\ev*{P}$ and $\ev*{\Delta P}$ are respectively the average power and power fluctuations, $\eta$ is the efficiency, and $\eta_\text{c}\equiv 1-\beta_\text{H}/\beta_\text{C}$ is the Carnot efficiency.
Such thermodynamic uncertainty relations imply, for example, that high efficiency can only be attained at the expense of low power or high power fluctuations. 
The thermodynamic uncertainty relation inequality \eqref{eq:tur_ineq} can be violated with quantum coherence~\cite{Agarwalla2018a,Brandner2018b,Krzysztof2018,Carollo2019,Liu2019,Kalaee2021,Rignon-Bret2021,Ehrlich2021,Gerry2022} and in PDHE~\cite{barato2016,proesmans2017,holubec2018,cangemi2020,menczel2021,lu2022}.
This  has motivated various generalised thermodynamic uncertainty relations~\cite{koyuk2019_jpmt,koyuk2019_prl,koyuk2020,VanVu2020,potanina2021,Hasegawa2021b}, in particular for time-symmetric driving~\cite{proesmans2017,timpanaro2019} and  slowly driven stochastic engines~\cite{miller2021,miller2021c}.

Despite their importance, thermodynamic uncertainty relations provide an incomplete picture of the trade-off: while high values of $\xi$ may appear more favorable, this does not give us any information on the individual objectives. Indeed, \cite{holubec2018,miller2021} have shown that high values of $\xi$ can be achieved in the limit of vanishing power, while often the goal is to operate at high power or efficiency. 

In this paper, we propose a framework to  optimize arbitrary trade-offs between power, efficiency, and power fluctuations in arbitrary PDHE described by Lindblad dynamics~\cite{gorini1976,lindblad1976,breuer2002,yamaguchi2017}; this framework enables the use of various optimization techniques, such as Pontryagin Minimum Principle \cite{kirk2004} or Reinforcement Learning (RL) \cite{sutton2018} to find Pareto-optimal cycles, i.e. those cycles where no objective can be further improved without sacrificing another one.
We then employ RL to fully optimize a quantum dot (QD) based heat engine~\cite{sothmann2015}. 
We characterize the Pareto front, i.e. the set of values $\{\ev*{P},\ev*{\Delta P}, \eta\}$ corresponding to Pareto-optimal cycles, and  evaluate the thermodynamic uncertainty relation ratio $\xi$ on such optimal cycles. Furthermore,
we derive analytical results for the Pareto front and $\xi$ in the fast~\cite{geva1992,schmiedl2008,cavina2018,erdman2019_njp,pekola2019,cangemi2021,cavina2021} and slow~\cite{salamon1983,esposito2010_prl,wang2011,avron2012,ludovico2016,cavina2017_prl,miller2019,scandi2019,Brandner2020,bhandari2020,abiuso2020_prl,abiuso2020_entropy,Frim2021,eglinton2022} driving regime, i.e. when the period of the cycle is respectively much shorter or much longer than the thermalization timescale of the system.

\paragraph*{Multi-objective optimization of quantum heat engines.}
\begin{figure}[!tb]
	\centering
	\includegraphics[width=0.99\columnwidth]{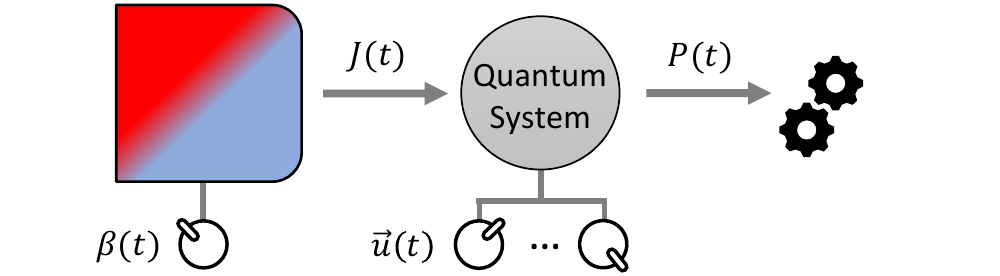}
	\caption{A quantum system (gray circle) is coupled to a thermal bath (left box) characterized by a controllable inverse temperature $\beta(t)$. The coupling produces a heat flux $J(t)$. Control parameters $\vec{u}(t)$ allow us to control the state of the system and the power $P(t)$ extracted from the system.}
	\label{fig:setup}
\end{figure}
We describe a PDHE as a quantum system coupled to a heat bath whose inverse temperature $\beta(t)$ can be tuned in time between two extremal values $\beta_\text{H}$ and $\beta_\text{C}$ with $\beta_\text{H} \leq \beta_\text{C}$ (Fig.~\ref{fig:setup}). The coupling produces a heat flux $J(t)$ from the bath to the quantum system. The PDHE is further controlled by time-dependent control parameters $\vec{u}(t)$ that allow exchanging work with the system, producing power $P(t)$. A thermodynamic cycle is then described by periodic functions $\beta(t)$ and $\vec{u}(t)$ with period $\tau$. This framework includes standard PDHE, in which the system is sequentially put in contact with two baths (by abruptly changing the values of $\beta(t)$) and cases where $\beta(t)$ varies smoothly in time~\cite{Brandner2015,Brandner2016,Brandner2020,miller2021,Frim2021,Miangolarra2021,Miangolarra2021b}.
We assume that the dynamics of the system  is described by a Markovian master equation, i.e that the reduced density matrix $\rho_t$ of the quantum system satisfies
\begin{equation}
\dot{\rho}_t = \mathcal{L}_{\vec{u}(t),\beta(t)}\rho_t,
\label{eq:master_eq}
\end{equation}
where $\mathcal{L}_{\vec{u}(t),\beta(t)}$ is the Lindbladian describing the evolution of the system \cite{breuer2002}.

We are interested in characterizing the performance of the PDHE computing the average power $\ev*{P}$, average irreversible entropy production $\ev*{\Sigma}$ and average power fluctuations $\ev*{\Delta P}$ in the asymptotic limit cycle
~\cite{Brandner2015,Brandner2016,miller2021}, i.e., in the limit of infinite repetitions of the cycle.
In such limit, $\rho_t$ becomes periodic with the same periodicity $\tau$ of the control (Sup. Mat.~\cite{supmat}).

Given the density matrix $\rho_t$, the calculation of $\ev*{P}$ and $ \ev*{\Sigma}$ can be performed using the standard approach first put forward in Ref.~\cite{alicki1979} (see Sup. Mat. ~\cite{supmat} for details).
Defining the time average $\ev{O}$ of an arbitrary quantity $O(t)$ as
$
    \ev{O} \equiv \tau^{-1}\int_0^\tau O(t)\, dt,
$
 we can calculate $\ev*{P}$ and  $\ev*{\Sigma}$ by averaging 
\begin{align}
    P(t) &= -\text{Tr}[\rho_t\dot{H}_{\vec{u}(t)}],
     &
     \Sigma(t) &=  -\text{Tr}[\dot{\rho}_t  H_{\vec{u}(t)}] \beta(t).
\label{eq:p_sigma}
\end{align}
Notice that we compute the entropy production  $\Sigma(t) \equiv -J(t)\beta(t)$ neglecting the entropy variation $\Delta S$ of the quantum system, since the periodicity of the state in the limiting cycle implies that $\Delta S=0$ after each repetition of the cycle.

The average fluctuations $\ev*{\Delta P}$, however, cannot be expressed as a time average of a function of the state $\rho_t$, since they involves a two-point correlation function. Indeed, from Ref.~\cite{miller2019} we can express them as
\begin{equation}
    \ev*{\Delta P} = \lim_{T\to\infty}  \frac{1}{T}\int_0^T dt \Tr[s_{t} \dot{H}_{\vec{u}(t)}],
    \label{eq:sigma_2}
\end{equation}
where we define
\begin{equation}
    s_{t} \equiv \int_0^{t} dt^\prime P(t,t^\prime)[ \dot{H}_{\vec{u}(t^\prime)}\rho_{t^\prime}] + \ev{w}_{t}\rho_{t} + h.c. \;.
    \label{eq:s_def}
\end{equation}
In Eq.~(\ref{eq:s_def}) $P(t,t^\prime) \equiv \overleftarrow{T}\exp[\int_{t^\prime}^{t}dt''\mathcal{L}_{\vec{u}(t''),\beta(t'')}]$ is the propagator, $\ev{w}_{t}\equiv-\int_0^{t}dt^\prime\Tr[\rho_{t^\prime}\dot{H}_{\vec{u}(t^\prime)}]$ is the total average work extracted between time $0$ and $t$, and $h.c.$ represents the complex conjugate of the right-hand-side.

Here, we overcome the difficulty of computing nested integrals and two-point correlation function in Eqs.~(\ref{eq:sigma_2}) and (\ref{eq:s_def}) by noticing that $s_t$ is a traceless Hermitian operator that satisfies
\begin{equation}
    \dot{s}_t = \mathcal{L}_{\vec{u}(t),\beta(t)} s_t + \{ \rho_t, \dot{H}_{\vec{u}(t)}\} - 2\text{Tr}[\rho_t\dot{H}_{\vec{u}(t)}]\rho_t,
    \label{eq:s_ode_main}
\end{equation}
and becomes periodic with period $\tau$ in the limiting cycle (Sup. Mat. \cite{supmat}). 

Therefore, by considering $(\rho_t, s_t)$ as an ``extended state'' satisfying the equations of motion in (\ref{eq:master_eq}) and (\ref{eq:s_ode_main}) in the time-interval $[0,\tau]$ with periodic boundary conditions, we can compute $\ev{P}$, $\ev{\Sigma}$ and $\ev*{\Delta P}$ as time-averages of $P(t)$, $\Sigma(t)$, and $\Delta P(t)$, where $P(t)$ and $\Sigma(t)$ are defined in Eq.~(\ref{eq:p_sigma}), and where
\begin{equation}
    \Delta P(t) \equiv \Tr[s_{t} \dot{H}_{\vec{u}(t)}].
    \label{eq:dp}
\end{equation}
Notice that these are now linear functionals of the extended state.

To identify Pareto-optimal cycles
we introduce the dimensionless figure of merit
\begin{equation}
\label{eq:F_def}
    \ev*{F} = a \frac{\ev*{P}}{P_\text{max}} - b\frac{\ev*{\Delta P}}{\Delta P(P_\text{max})} - c \frac{\ev*{\Sigma}}{\Sigma(P_\text{max})},
\end{equation}
where $a,b,c\geq 0$ are three scalar weights, satisfying $a+b+c=1$, that determine how much we are interested in each of the three objectives, and  $P_\text{max}$, $\Delta P (P_\text{max})$ and $\Sigma(P_\text{max})$ are respectively the average power, fluctuations and entropy production of the cycle that maximizes the power.
Notice that, given the relation between entropy production and efficiency, cycles that are Pareto-optimal for $\{\ev{P},\ev{\Delta P},\eta\}$, are also Pareto-optimal for $\{\ev{P},\ev{\Delta P},\ev*{\Sigma}\}$ (Sup. Mat.~\cite{supmat}). 
The positive sign in front of $\ev*{{P}}$ in Eq.~(\ref{eq:F_def}) ensures that we are maximizing the power, while the negative sign in front of $\ev*{\Delta {P}}$ and $\ev*{\Sigma}$ ensures that we are minimizing power fluctuations and the entropy production. 
Interestingly, if convex, it has been shown that the full Pareto front can be identified repeating the optimization of $\ev*{F}$ for many values of $a$, $b$, and $c$ \cite{seoane2016, solon2018}.

Since $\ev{F}$ is a linear combination of the average thermodynamic quantities, using Eqs.~(\ref{eq:p_sigma}) and (\ref{eq:dp}) we can express $\ev{F}$ as a time integral of a function of the extended state $(\rho_t,s_t)$ and of the controls $\vec{u}(t)$ and $\beta(t)$:
\begin{equation}
    \ev{F} = \int_0^\tau G\left(\rho_t, s_t, \vec{u}(t), \beta(t)\right) dt,
\end{equation}
where $G(\rho_t, s_t, \vec{u}(t), \beta(t))$ is a suitable function.
The optimization of $\ev*{F}$ in this form is precisely the type of problem that can be readily tackled using optimization techniques such as Pontryagin Minimum Principle~\cite{kirk2004} or RL~\cite{sutton2018}. In this manuscript, we employ the latter.

\paragraph*{QD heat engine.}
In the following, we compute Pareto-optimal cycles in a minimal heat engine consisting of a two-level system coupled to a Fermionic bath with flat density of states. This represents a model of a single-level QD 
\cite{geva1992, esposito2010_pre, sothmann2015}.
The Hamiltonian reads
\begin{equation}
    H_{u(t)} = u(t) \frac{E_0}{2} \sigma_z,
\end{equation}
where $u(t)$ is our single control parameter, $E_0$ is a fixed energy scale and $\sigma_z$ is a Pauli matrix. Denoting with $\ket{1}$ the excited state of $H_{u(t)}$, and defining $p_t\equiv \mel*{1}{\rho_t}{1}$ as the probability of being in the excited state, the Lindblad equation (\ref{eq:master_eq}) becomes 
$
    \dot{p}_t = -\gamma(p_t - \pi_{\vec{u}(t),\beta(t)}),
$
where $\gamma^{-1}$ is the thermalization timescale arising from the coupling between  system and  bath, and $\pi_{\vec{u}(t),\beta(t)} = f(\beta(t)E_0u(t))$ is the excited level population of the instantaneous Gibbs state, with $f(x) \equiv (1+ e^x)^{-1}$~\cite{erdman2019_njp}.

\paragraph*{Optimal cycles with RL and analytical results.}
We optimize $\ev*{F}$ of the QD heat engine using three different tools: RL, analytics in the fast-driving regime, and analytics in the slow-driving regime. 

The RL-based method allows us to numerically optimize $\ev*{F}$ without making any approximations on the dynamics, exploring
all possible (time-discretized) 
time dependent controls $\beta(t)$ and $u(t)$ subject to the constraints $\beta(t)\in[\beta_\text{H},\beta_\text{C}]$ and $u(t)\in [u_\text{min},u_\text{max}]$ (thus beyond fixed structures such as Otto cycles), and identifying automatically  also the optimal period. The RL method, based on the soft actor-critic algorithm~\cite{haarnoja2018_arxiv_sac,haarnoja2018_arxiv_walk,haarnoja2018_pmlr} and generalized from  \cite{erdman2022,erdman2022_arxiv}, additionally includes the crucial impact of power fluctuations, and identifies Pareto-optimal cycles (Sup. Mat.~\cite{supmat} for technical details and for benefits of using RL).
Machine learning methods have been employed for other quantum thermodynamic~\cite{sgroi2021,khait2021, ashida2021} and quantum control \cite{bukov2018,fosel2018,an2019,niu2019,zhang2019,dalgaard2020,mackeprang2020,schafer2020,sweke2020,schafer2021,porotti2021,metz2022,coopmans2022}
 tasks.

The fast-driving regime assumes that $\tau \ll \gamma^{-1}$.
Interestingly, without any assumption on the driving speed, we show~\cite{supmat} that any trade-off between power and entropy production ($b=0$ in Eq.~\eqref{eq:F_def}) in the QD engine is maximized by Otto cycles in the fast-driving regime, i.e. switching between two values of $\beta(t)$ and $u(t)$ ``as fast as possible''~\cite{erdman2019_njp,cavina2021}. We thus expect such ``fast-Otto cycles'' to be nearly optimal in the high power or efficiency regime.    Moreover, we derive analytical expressions to compute and optimize $\{\ev{P},\ev{\Delta P},\ev*{\Sigma}\}$ efficiently in arbitrary systems in the fast driving regime~\cite{supmat}.

The slow-driving regime corresponds to the opposite limit,  i.e. $\tau \gg \gamma^{-1}$.  Since entropy production and power fluctuations can be minimized by considering quasi-static cycles~(see e.g. \cite{holubec2018,miller2021}), we expect this regime to be nearly optimal in the low power regime, i.e. for low values of $a$ in Eq.~\eqref{eq:F_def}. 
To make analytical progress in this regime, we  maximize Eq. \eqref{eq:F_def} assuming a finite-time Carnot cycle (c.f.  Sup. Mat.~\cite{supmat}). The obtained results naturally generalize previous considerations for low-dissipation engines~\cite{esposito2010_prl,esposito2010_pre,Holubec2015,Holubec2016,abiuso2018,Ma2018,abiuso2020_prl} to account for the role of fluctuations (see also \cite{Miller2020a}). The main technical tool  is the geometric concept of ``thermodynamic length''\cite{salamon1983,abiuso2020_entropy} which yields the-first order correction in $(\gamma\tau)^{-1}$ from the quasi-static limit.

\begin{figure}[!tb]
	\centering
	\includegraphics[width=0.99\columnwidth]{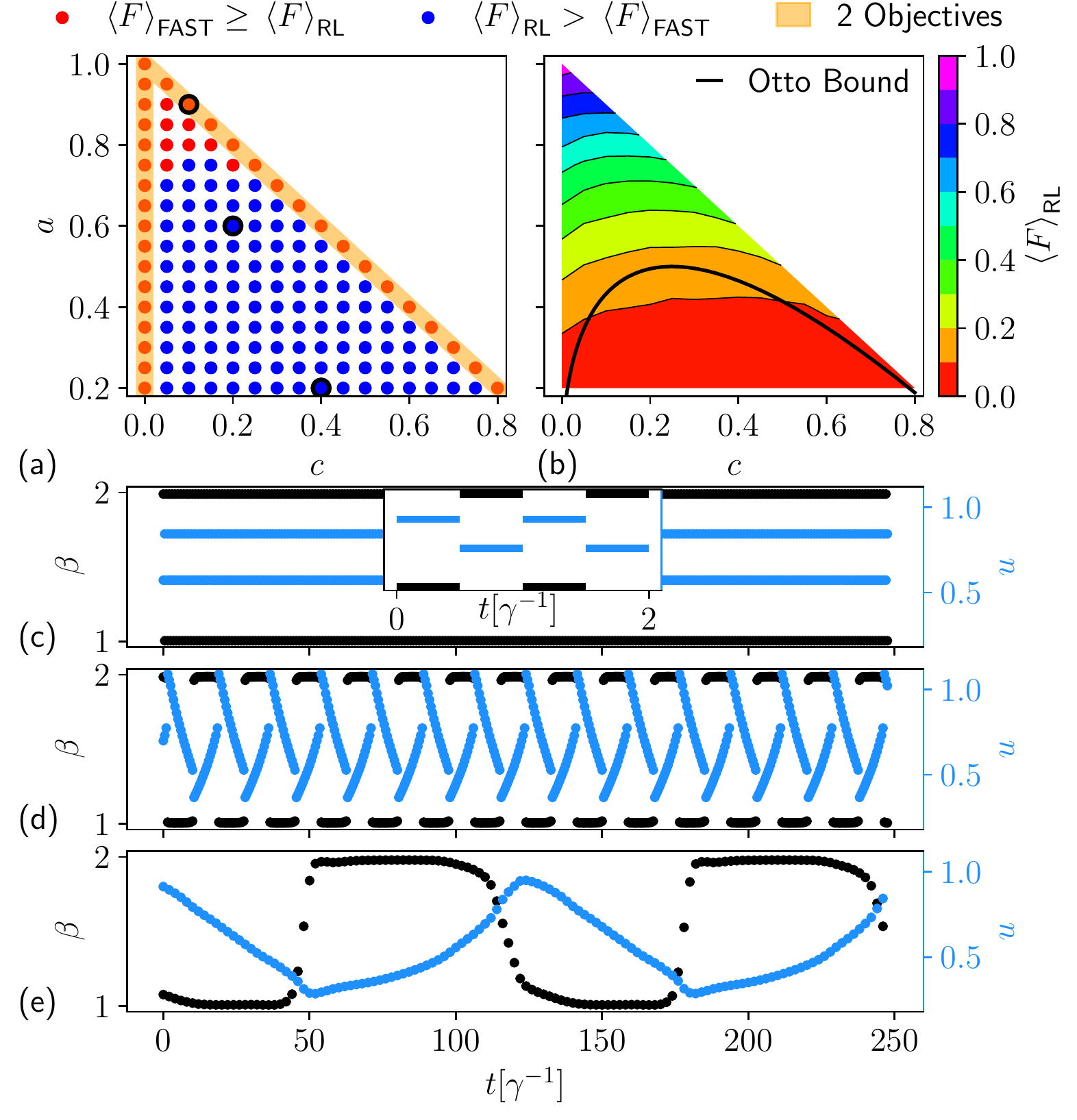}
	\caption{Optimization of $\ev*{F}$ at different values of $a$ and $c$, with $b=1-a-c$, for a QD-based PDHE. Each dot in panel (a) displays, as a function of $c$ and $a$, whether $\ev*{F}_\text{RL}>\ev{F}_\text{FAST}$ (blue dots) or not (red dots). Points with $a\sim 0$ are not displayed since, in such a regime, optimal cycles become infinitely long (to minimize entropy production and fluctuations) and the RL method does not converge reliably \cite{supmat}. (b): contour plot of $\ev*{F}_\text{RL}$, as a function of $c$ and $a$, using the data-points of (a). The black line represents the curve below which $\ev*{F}_\text{FAST} = 0$. (c,d,e): cycles, described by piece-wise constant values of $\beta$ (black dots) and $u$ (blue dots) as a function of $t$, identified at the three values of $a$ and $c$ highlighted in black in panel (a) (respectively from top to bottom). The inset in panel (c) represents a zoom into the corresponding cycle, which is a fast-Otto cycle. Parameters: $\beta_\text{C} = 2$, $\beta_\text{H}=1$, $u_\text{min}=0.2$,
$u_\text{max}=1.1$ and $E_0=2.5$.}
	\label{fig:fig_merit}
\end{figure}
We now present the results. Each point in Fig.~\ref{fig:fig_merit}a corresponds to a separate optimization of $\ev*{F}$ with weights $c$ and $a$ displayed on the x-y axis. Since $b=1-a-c$, points lying on the sides of the triangle (highlighted in yellow) correspond to optimizing the trade-off between 2 objectives, whereas points inside the triangle take all 3 objectives into account. 
Denoting the figure of merit optimized with RL and with fast-Otto cycles with $\ev*{F}_\text{RL}$ and $\ev*{F}_\text{FAST}$,
in Fig.~\ref{fig:fig_merit}a we show blue (red) dots  when $\ev{F}_\text{RL} > \ev{F}_\text{FAST}$ ($\ev{F}_\text{RL} \leq \ev{F}_\text{FAST}$), while Fig.~\ref{fig:fig_merit}b is a contour plot of $\ev{F}_\text{RL}$. As expected, there are red dots when $b=0$ (along the hypotenuse), but it turns out that fast-Otto cycles are optimal also when~$c=0$. However, as soon as all 3 weights are finite, the optimal cycles identified with RL change abruptly and outperform fast-Otto cycles. 
Furthermore, we notice that while $\ev{F}_\text{RL}$ is positive for all values of the weights, $\ev{F}_\text{FAST}=0$ below the black curve shown in Fig.~\ref{fig:fig_merit}b (see Sup. Mat.~\cite{supmat} for its analytic expression). 

To visualize the changes in protocol space, in Figs.~\ref{fig:fig_merit}(c,d,e) we show the cycles identified with RL at the three different values of the weights highlighted by a black circle in Fig.~\ref{fig:fig_merit}a (respectively from top to bottom). Since RL identifies piece-wise constant controls, the cycle is displayed as dots corresponding to the value of $\beta(t)$ (black dots) and $u(t)$ (blue dots) at each small time-step. First, we notice
that the inverse temperature abruptly switches between $\beta_\text{H}$ and $\beta_\text{C}$ for all values of the weights, so that in this engine no gain arises when smoothly varying the temperature.  
As expected, 
the cycle identified by RL in 
Fig.~\ref{fig:fig_merit}c, corresponding to the black point on the hypotenuse in Fig.~\ref{fig:fig_merit}a,
is a fast-Otto cycle (a ``zoom'' in a short time interval is shown in the inset). However, moving down in weight space to the black dot at $a=0.6$ and $c=0.2$, we see that the corresponding cycle (Fig.~\ref{fig:fig_merit}d) now displays a finite period, with  linear modulations of $u(t)$ at fixed temperatures, and a discontinuity of $u(t)$ when switching between $\beta_\text{H}$ and $\beta_\text{C}$. The cycle in Fig.~\ref{fig:fig_merit}e, corresponding to the lowest black dot at $a=0.2$ and $c=0.4$, displays an extremely long period $\tau \approx 125 \gamma^{-1}$, which is far in the slow-driving regime. 
Optimal cycles, therefore, interpolate between the fast and the slow-driving regimes as we move in weight space (Fig.~\ref{fig:fig_merit}a) from the sides to the lower and central region (i.e. switching from 2 to 3 objectives). 

\begin{figure}[!tb]
	\centering
	\includegraphics[width=0.99\columnwidth]{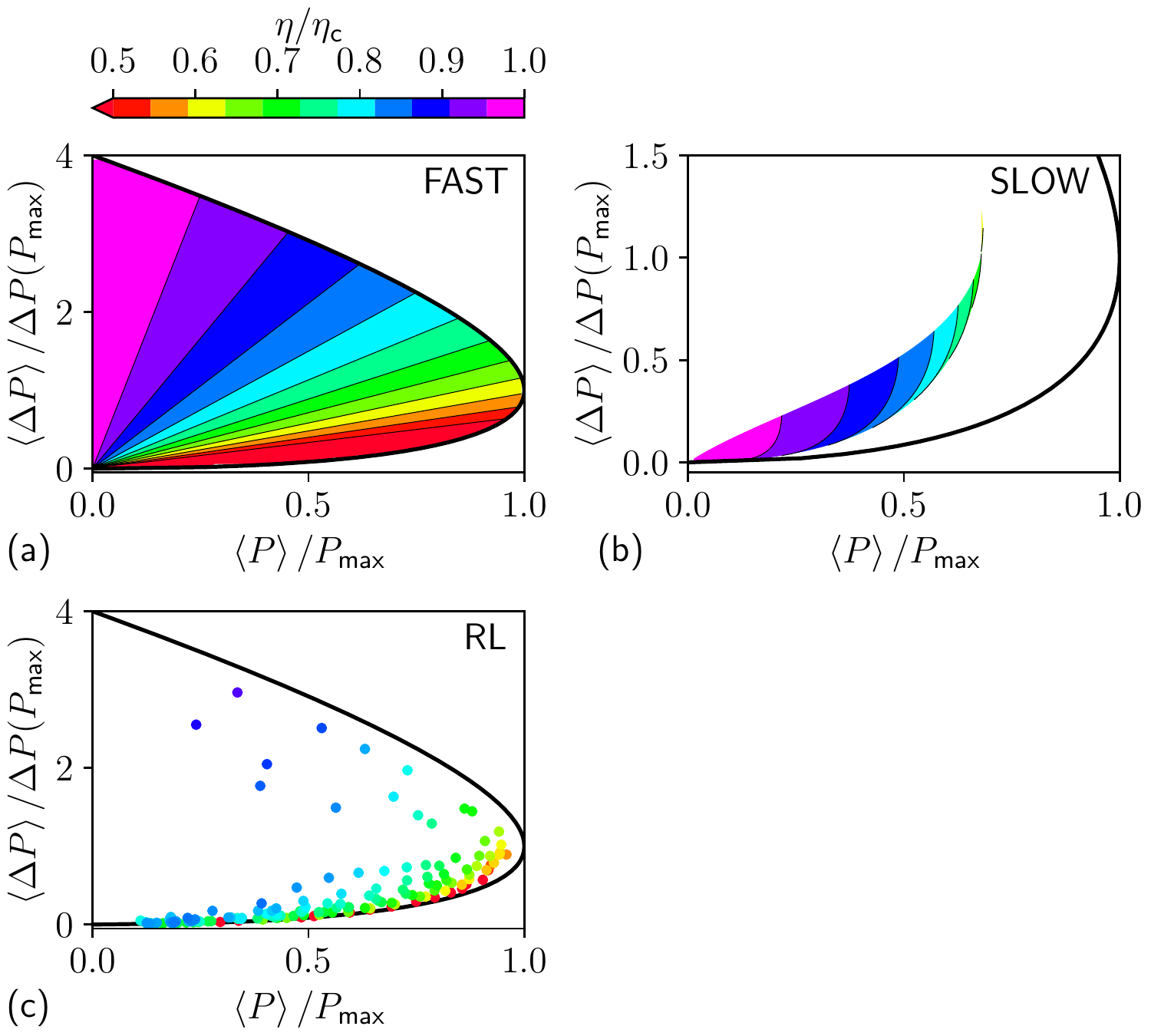}
	\caption{Pareto-front found optimizing $\ev*{F}$ with fast-Otto cycles in the limit of small temperature differences [panel (a)], optimizing $\ev*{F}$ in the slow-driving regime [panel (b)], and numerically using RL [panel (c)]. The system parameters are as in Fig.~\ref{fig:fig_merit}. All panel display $\ev*{\Delta {P}}/\Delta P(P_\text{max})$ as a function of $\ev*{{P}}/P_\text{max}$ (x-axis) and of $\eta/\eta_\text{c}$ (color). The black curve represents the outer border of the Pareto-front derived analytically in Sup. Mat.~\cite{supmat}.}
	\label{fig:full_pareto}
\end{figure}
In Fig.~\ref{fig:full_pareto} we display the Pareto-front, i.e. we plot the value of ${P}/P_\text{max}$, $\Delta {P}/\Delta P(P_\text{max})$, and of the efficiency $\eta/\eta_\text{c}$, found maximizing $\ev*{F}$ for various values of the weights. Figure~\ref{fig:full_pareto}a is derived in the fast-driving regime assuming a small temperature difference, while Fig.~\ref{fig:full_pareto}b is derived in the slow-driving regime. The RL results, shown in Fig.~\ref{fig:full_pareto}c, correspond to the points in Fig.~\ref{fig:fig_merit}a.
First, we notice that, by definition of the Pareto front, the ``outer border'' corresponds to points where we only maximize the trade-off between the two objectives $\ev*{P}$ and $\ev*{\Delta {P}}$. Since these points are optimized by fast-Otto cycles, the black border of Fig.~\ref{fig:full_pareto}a, also shown in Figs.~\ref{fig:full_pareto}b, \ref{fig:full_pareto}c, is exact
and given by (see Sup. Mat. \cite{supmat} for details)
\begin{equation}
    \frac{\ev*{P}}{P_\text{max}} = 2\sqrt{\frac{\ev*{\Delta P}}{\Delta P(P_\text{max})}}- \frac{\ev*{\Delta P}}{\Delta P(P_\text{max})}.
    \label{eq:pareto_analytic}
\end{equation}
Moreover, in this setup, we can establish an exact mapping between the performance of a SSHE and of our PDHE operated with fast-Otto cycles  (cf. Sup. Mat.~\cite{supmat}). Since SSHE satisfy Eq.~(\ref{eq:tur_ineq}), also fast-Otto cycles have $\xi\leq 1$.
Furthermore, for small temperature differences,~$\xi=1$. This allows us to fully determine the internal part of the Pareto front in the fast-driving regime using the thermodynamic uncertainty relations, i.e. ${P}/P_\text{max} = (\Delta {P}/\Delta P (P_\text{max}))(\eta_\text{c}-\eta)/\eta$. Indeed, the linear contour lines in Fig.~\ref{fig:full_pareto}a stem from the linearity between ${P}$ and $\Delta {P}$, the angular coefficient being determined by the efficiency. 

Comparing Figs.~\ref{fig:full_pareto}(a,b), we see where the fast and slow-driving regimes are optimal. 
As expected, the slow-driving Pareto front cannot reach the black border, especially in the high-power area, where fast-Otto cycles are optimal. However, in the low power and low fluctuation regime, cycles in the slow-driving substantially outperform fast-Otto cycles by delivering a higher efficiency (purple and blue regions in Fig.~\ref{fig:full_pareto}b).

\begin{figure}[!tb]
	\centering
	\includegraphics[width=0.99\columnwidth]{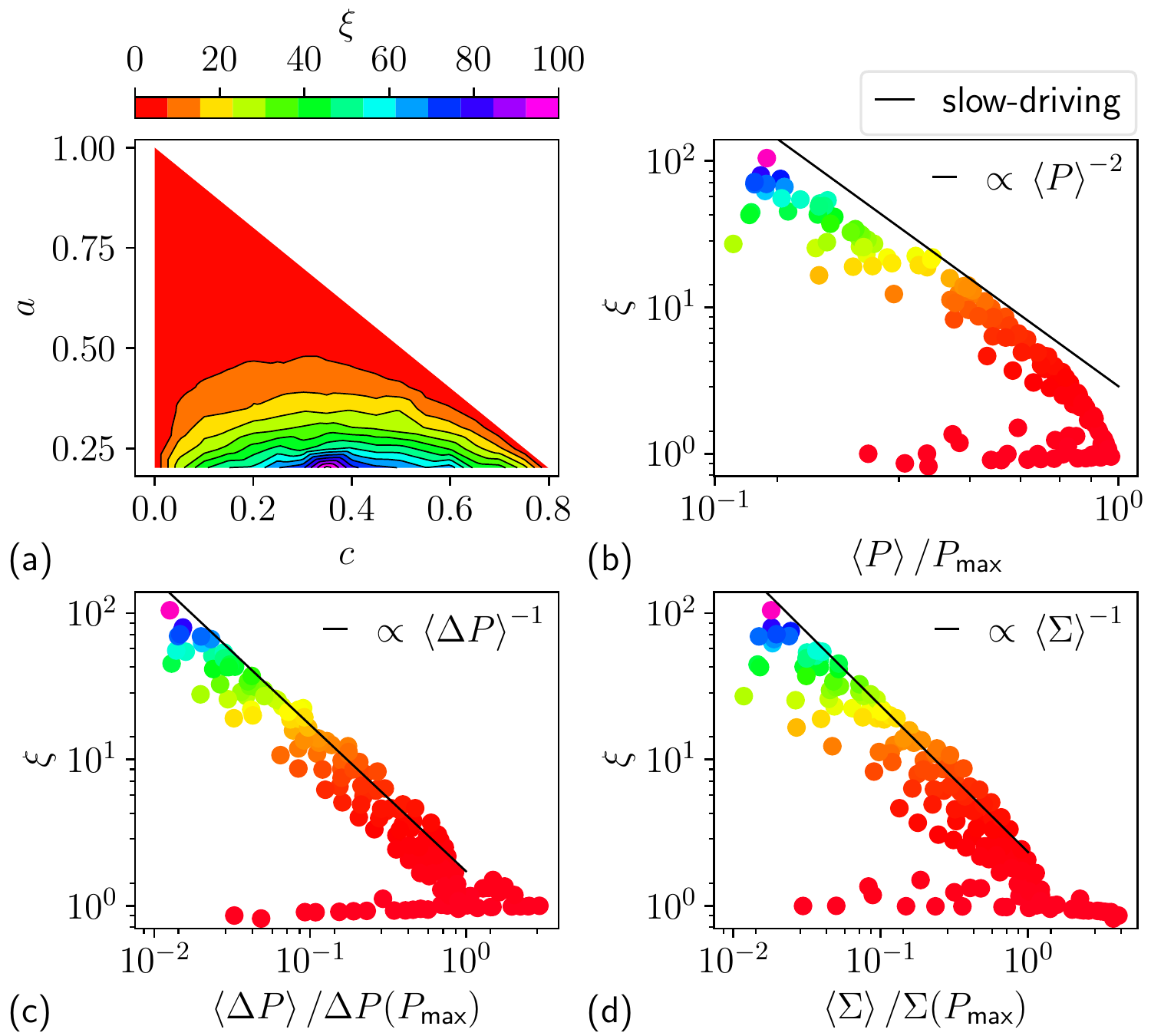}
	\caption{(a): contour plot of the SSHE thermodynamic uncertainty relationship ratio $\xi$ as a function of $c$ and $a$. (b,c,d): log-log plot of $\xi$, color mapped as in panel (a), as a function of $\ev*{{P}}/P_\text{max}$, $\ev*{\Delta {P}}/\Delta P( P_\text{max})$ and $\ev*{\Sigma}/\Sigma(P_\text{max})$, respectively. 
	Every point corresponds to the same RL optimization performed in Fig.~\ref{fig:fig_merit}. The black line is the behavior of $\xi$ derived analytically  in Sup. Mat.~\cite{supmat} in the slow-driving regime for small values of $\ev*{{P}}$, $\ev*{\Delta {P}}$ and $\ev*{\Sigma}$.}
	\label{fig:tur}
\end{figure}
Interestingly, the RL points in Fig.~\ref{fig:full_pareto}c capture the best features of both regimes. RL can describe the high-power and low fluctuation regime displaying both red and blue/green dots near the lower border. The red dots are fast-Otto cycles that are optimal exactly along the border but deliver a low efficiency. The blue/green dots instead are finite-time cycles that deliver a much higher efficiency by sacrificing a very small amount of power and fluctuations. This dramatic enhancement of the efficiency as we depart from the lower border is another signature of the abrupt change in optimal cycles.

\emph{Violation of thermodynamic uncertainty relation}. At last, we analyze the behavior of the thermodynamic uncertainty relation ratio $\xi$, which represents a relevant quantity combining the three objectives, computing it on Pareto-optimal cycles (recall that $\xi\leq 1$ for classical stochastic SSHE but  PDHE can violate this bound~\cite{barato2016,proesmans2017,holubec2018,cangemi2020,menczel2021,lu2022}).
In Fig.~\ref{fig:tur}a we show a contour plot of $\xi$, computed with RL, as a function of~$a$ and~$c$. Because of the mapping between SSHE and fast-Otto cycles \cite{supmat}, we have $\xi = 1$ along the sides of the triangle, where only 2 objectives are optimized. However, this mapping breaks down for finite-time cycles, allowing us to observe a strong increase of $\xi$ 
in the green/purple region in Fig.~\ref{fig:tur}a. As shown in Fig.~\ref{fig:fig_merit}, this region corresponds to long cycles operated in the slow-driving regime, where violations of thermodynamic uncertainty relations had already been reported~\cite{holubec2018,miller2021}. In Figs.~\ref{fig:tur}(b,c,d) we show a log-log plot of $\xi$ respectively as a function of ${P}/P_\text{max}$, $\Delta {P}/\Delta P(P_\text{max})$, and ${\Sigma}/\Sigma(P_\text{max})$ with the same color-map as in Fig.~\ref{fig:tur}a. We see that $\xi$ diverges in the limit of low power, low fluctuations, and low entropy production as a power law. 
Indeed, using the slow-driving approximation, we analytically prove that $\xi$ diverges as $\ev*{{P}}^{-2}$, $\ev*{\Delta {P}}^{-1}$, and $\ev*{\Sigma}^{-1}$. Such relations, plotted as black lines, nicely agree with our RL results.

\paragraph*{Conclusions.}
We introduced a general framework to identify Pareto-optimal cycles between power, efficiency, and power fluctuations in quantum or classical stochastic heat engines, paving the way for their systematic optimization using optimal control techniques such as the Pontryagin Minimum Principle \cite{kirk2004} or Reinforcement Learning \cite{sutton2018}. As opposed to previous literature reviewed above, we account for the crucial impact of power fluctuations.
We then employed RL to optimize a quantum dot-based heat engine, solving its exact finite-time and out-of-equilibrium dynamics, providing us with new physical insights.
We observe an abrupt change in Pareto-optimal cycles when switching from the optimization of $2$ objectives, where Otto cycles in the fast-driving regime are optimal, to 3 objectives, where the optimal cycles have a finite period.
This feature, which shares analogies with the phase transition in protocol space observed in Ref.~\cite{solon2018}, corresponds to a large enhancement of one of the objectives while only slightly decreasing the other ones. 
Furthermore, we find an exact mapping between Otto cycles in the fast-driving regime and SSHEs, implying that a violation of the thermodynamic uncertainty relation ratio $\xi$ in Eq.~(\ref{eq:tur_ineq}) requires the optimization of all $3$ objectives.
We then find that $\xi$ becomes arbitrarily large in the slow-driving regime. Cycles found with RL display the best features analytically identified in the fast and slow driving regimes.

\paragraph*{Acknowledgments.}
FN gratefully acknowledges funding by the BMBF (Berlin Institute for the Foundations of Learning and Data -- BIFOLD), the European Research Commission (ERC CoG 772230) and the Berlin Mathematics Center MATH+ (AA1-6, AA2-8). PAE gratefully acknowledges funding by the Berlin Mathematics Center MATH+ (AA1-6, AA2-18).  AR and MPL acknowledge funding from
Swiss National Science Foundation through an
Ambizione Grant No. PZ00P2-186067.
PA is supported by “la Caixa” Foundation (ID 100010434, Grant No. LCF/BQ/DI19/11730023), and by the Government of Spain (FIS2020-TRANQI and Severo Ochoa CEX2019-000910-S), Fundacio Cellex, Fundacio Mir-Puig, Generalitat de Catalunya (CERCA, AGAUR SGR 1381).

%

\pagebreak
\ 
\newpage
\widetext
\begin{center}
\textbf{\large Supplemental Material for ``Pareto-optimal cycles for power, efficiency and fluctuations of quantum heat engines using reinforcement learning''}
\end{center}
\setcounter{equation}{0}
\setcounter{figure}{0}
\setcounter{table}{0}
\setcounter{page}{1}
\setcounter{section}{0}
\makeatletter
\renewcommand{\theequation}{S\arabic{equation}}
\renewcommand{\thefigure}{S\arabic{figure}}

\section{General Framework and Reinforcement Learning approach}
In this section we describe a general framework to compute the power, entropy production and power fluctuations of a quantum thermal machine. We then show how it allows us to use RL to optimize the trade-off between these three objectives.
Assuming that the dynamics is Markovian, the reduced  state evolves according to Eq.~(\ref{eq:master_eq}) of the main text, i.e.
\begin{equation}
    \dot{\rho}_t = \mathcal{L}_{\vec{u}(t),\beta(t)} \rho_t.
    \label{eq:lindblad}
\end{equation}
As usual, we employ the first law to split the internal energy $U(t)=\Tr[\rho_t H_{\vec{u}(t)}]$ of the system into absorbed heat and output work \cite{alicki1979}:
\begin{align}
    dU=dQ-dW\;, \quad dQ=\Tr[d\rho_t H_{\vec{u}(t)}]\;,\quad dW=-\Tr[\rho_t dH_{\vec{u}(t)}]\;.
\end{align}
The instantaneous power \emph{output} is thus given by Eq.~(\ref{eq:p_sigma}) of the main text, i.e.
\begin{align}
P(t)=-\Tr[\rho_t\dot{H}_{\vec{u}(t)}] 
\end{align}
and, according to Clausius theorem for cycles (in which we set $\Delta S=0$ of the quantum system since we consider a periodic driving in the limiting cycle), the instantaneous irreversible entropy production is given by Eq.~(\ref{eq:p_sigma}) of the main text, i.e.
\begin{align}
    \Sigma(t)=-\beta(t)\Tr[\dot{\rho}_t H_{\vec{u}(t)}]\;.
\end{align}
For what concerns power fluctuations, let $\langle\sigma^2\rangle_t$ be the work fluctuations between time $0$ at $t$. Using Eq.~(2) of Ref.~\cite{miller2019}, we can write the fluctuations as
\begin{equation}
    \langle\sigma^2\rangle_t = \int_0^t dt_1 \Tr[\dot{H}_{\vec{u}(t_1)} s_{t_1}],
    \label{eq:sigma_2_sm}
\end{equation}
where we introduce
\begin{equation}
    s_{t_1} \equiv \int_0^{t_1} dt_2 P(t_1,t_2)[ \dot{H}_{\vec{u}(t_2)}\rho_{t_2}] + \ev{w}_{t_1}\rho_{t_1} + h.c. \;.
    \label{eq:s_def_sm}
\end{equation}
Here $P(t_1,t_2) \equiv \overleftarrow{T}\exp[\int_{t_2}^{t_1}d\tau\mathcal{L}_{\vec{u}(\tau),\beta(\tau)}]$ is the propagator, $\ev{w}_{t_1}\equiv-\int_0^{t_1}dt_2\Tr[\rho_{t_2}\dot{H}_{\vec{u}(t_2)}]$ is the total average work extracted between time $0$ and $t_1$, $h.c.$ represents the complex conjugate of the whole right-hand-side, and in deriving Eq.~(\ref{eq:sigma_2_sm}) we used that $[P(t_1,t_2)O]^\dagger = P(t_1,t_2)O^\dagger$, which stems from $[\mathcal{L}_{\vec{u}(\tau),\beta(\tau)}O]^\dagger = \mathcal{L}_{\vec{u}(\tau),\beta(\tau)}O^\dagger$. Taking the time-derivative of $s_t$, we find that it satisfies the equation of motion given in Eq.~(\ref{eq:s_ode_main}) of the main text, i.e.
\begin{equation}
    \dot{s}_t = \mathcal{L}_{\vec{u}(t),\beta(t)} s_t + \{ \rho, \dot{H}_{\vec{u}(t)}\} - 2\Tr[\rho_t\dot{H}_{\vec{u}(t)}]\rho_t,
    \label{eq:s_ode}
\end{equation}
where $\{\cdot,\cdot\}$ denotes the anti-commutator.
Using the definition of $s_t$ given in Eq.~(\ref{eq:s_def_sm}), we see that $s_t$ is an Hermitian operator, and that $\Tr[s_t] = 0$. This last property stems from the fact that $P(t_1,t_2)[\cdot]$ is trace-preserving.

\subsection{``Extended state'' and formulation of the optimization problem}
\label{sec:extended_state}
We now define $(\rho_t,s_t)$ as an ``extended state''. We have just shown that the equation of motion of the extended state is a set of first-order differential equations [see Eqs.~(\ref{eq:lindblad}) and (\ref{eq:s_ode})]. Furthermore, the average power $\ev{P} \equiv \ev{w}_\tau/\tau$, the average fluctuations $\ev{\Delta P} \equiv \langle\sigma^2\rangle_\tau/\tau$ and the average entropy production $\ev{\Sigma}\equiv (1/\tau)\int_0^\tau \Sigma(t)dt$ can be expressed as a time-average of the extended state and of the controls $\vec{u}(t)$ and $\beta(t)$, i.e.
\begin{equation}
\begin{aligned}
\label{eqapp:P_dP_Sigma}
    \ev{P} &= -\frac{1}{\tau} \int_0^\tau dt \Tr[\rho_t\dot{H}_{\vec{u}(t)}]=\frac{1}{\tau} \int_0^\tau dt \Tr[\dot{\rho}_t H_{\vec{u}(t)}], \\
    \ev{\Delta P} &= \frac{1}{\tau} \int_0^\tau dt \Tr[s_t \dot{H}_{\vec{u}(t)}], \\
    \ev{\Sigma} &= -\frac{1}{\tau} \int_0^\tau dt \Tr[\dot{\rho}_t H_{\vec{u}(t)}]\beta(t).
\end{aligned}
\end{equation}
Therefore, also the figure of merit $\ev{F}$ in Eq.~(\ref{eq:F_def}) of the main text can be expressed as a time integral of a function of the extended state $(\rho_t, s_t)$ and of the controls $\vec{u}(t)$ and $\beta(t)$, i.e. 
\begin{equation}
    \ev{F} = \int_0^\tau G\left(\rho_t, s_t, \vec{u}(t), \beta(t)\right) dt,
\end{equation}
where $G(\rho_t, s_t, \vec{u}(t), \beta(t))$ is a suitable function.
This is precisely the type of problem that can readily tackled using optimal control techniques such as Pontryagin Minimum Principle \cite{kirk2004} or Reinforcement Learning \cite{sutton2018}.

As a final remark, one can study heat engines in the limiting cycle imposing periodic boundary conditions for the extended state, i.e. $(\rho_0, s_0) = (\rho_\tau, s_\tau) $. Indeed, we are interested in studying thermodynamic cycles, i.e. when both the control and the quantum state $\rho_t$ are periodic with period $\tau$. Assuming that the Lindbladian $\mathcal{L}_{\vec{u}(t),\beta(t)}$ has a single fixed point, it can be shown that a periodic control eventually drives the state $\rho_t$ towards the limiting cycle solution, where $\rho_t$ becomes periodic with the same period $\tau$ of the driving \cite{menczel2019_jpmt}. 
Furthermore, the equation of motion of $s_t$ has the same homogeneous part ($\mathcal{L}_{\vec{u}(t),\beta(t)}[s_t]$) as $\rho_t$, and the non-homogeneous term becomes periodic once $\rho_t$ reaches the limiting cycle. Therefore, also $s_t$ naturally reaches a limiting cycle solution where it is periodic with the same period $\tau$.

\subsection{Optimizing the entropy production instead of the efficiency}
Here we discuss the relation between minimizing the entropy production or maximizing the efficiency. We generalize the discussion of Ref.~\cite{erdman2022_arxiv}, where power fluctuations are not considered.
We start by noticing that we can express the efficiency of a heat engine in terms of the average power and entropy production, i.e.
\begin{equation}
	\eta = \eta_\text{c}\, [1+\ev{\Sigma}/(\beta_\text{C}\ev*{P}) ]^{-1}.
	\label{eq:eta}
\end{equation}
We show that, thanks to this dependence of $\eta$ on $\ev*{P}$ and $\ev*{\Sigma}$, optimizing a trade-off between high power, low fluctuations and high efficiency yields all the Pareto optimal trade-offs between high power, low fluctuations, and low entropy-production up to a change of the weights $(a,b,c)$.

First, we provide a non-rigorous argument as follows: notice that a point $\{\ev{P},\ev{\Delta P},\ev{\Sigma}\}$ belongs to the Pareto front iff there exists no cycle outperforming it in all 3 quantities, i.e. larger power, smaller fluctuations and smaller entropy production. Due to the functional dependence of $\eta(\ev{\Sigma},\ev{P})$~\eqref{eq:eta}, the corresponding point $\{\ev{P},\ev{\Delta P},\eta(\ev{\Sigma},\ev{P})\}$ will belong to the power-fluctuations-efficiency Pareto front. In fact, if the point $\{\ev{P},\ev{\Delta P},\eta(\ev{\Sigma},\ev{P})\}$ is not on the Pareto front, then a cycle $\{\ev{P}',\ev{\Delta P}',\eta'\}$ exists such that one of the following statements holds
\begin{align}
\text{(Case 1)}\quad
    \ev{P}'>\ev{P}\;, \quad \ev{\Delta P}'=\ev{\Delta P}\;, \quad \eta'=\eta(\ev{\Sigma},\ev{P})\;;
\end{align}
\begin{align}
\text{(Case 2)}\quad
    \ev{P}'=\ev{P}\;, \quad \ev{\Delta P}'<\ev{\Delta P}\;, \quad \eta'=\eta(\ev{\Sigma},\ev{P})\;;
\end{align}
\begin{align}
\text{(Case 3)}\quad
    \ev{P}'=\ev{P}\;, \quad \ev{\Delta P}'=\ev{\Delta P}\;, \quad \eta'>\eta(\ev{\Sigma},\ev{P})\;.
\end{align}
Notice that in (Case 2) the corresponding $\{\ev{P}',\ev{\Delta P}',\ev{\Sigma}(\ev{P}',\eta')\}\equiv \{\ev{P},\ev{\Delta P}',\ev{\Sigma}\} $ clearly  violates the power-fluctuations-entropy production Pareto front, i.e. outperforms the point $\{\ev{P},\ev{\Delta P},\ev{\Sigma}\}$ because of smaller fluctuations. Similarly, by inverting~\eqref{eq:eta}
\begin{equation}
    \Sigma(P,\eta)= \frac{\eta_\text{c}-\eta}{\eta} \beta_\text{C}P
    \label{eq:sigma_p_eta}
\end{equation}
we see that in (Case 3) the corresponding point violates the Pareto front via a reduction of $\ev{\Sigma}$. Finally (Case 1) is non trivial, but it is intuitive (and empirically verified) that in power-fluctuations-efficiency trade-offs it is always possible to reduce the power in favor of efficiency and fluctuations, (typically by slowing down the whole cycle, i.e. increasing $\tau$). This means that the occurrence of (Case 1) induces the existence of (Case 2) and (Case 3), closing the argument.

Besides the above intuitive argument, to prove such equivalence mathematically, we can prove that the cycles that maximize 
\begin{equation}
	\ev*{G} \equiv a \frac{\ev*{P}}{P_\text{max}} -b\frac{\ev*{\Delta P}}{\Delta P(P_\text{max})} +c\, \frac{\eta}{\eta_\text{c}}
\end{equation}
for some values of the weights $a\geq 0$, $b\geq 0$, $c\geq 0$ such that  $a+b+c=1$, also maximize the figure of merit in Eq.~(\ref{eq:F_def}) of the main text for some (possibly different) non-negative values of the weights summing to one. To simplify the proof and the notation, we consider the following two functions
\begin{equation}
\begin{aligned}
	F(a,b,c;\theta) &= a P(\theta) -b\Delta P(\theta) - c \Sigma(P(\theta),\eta(\theta)),\\
	G(a,b,c;\theta) &= a P(\theta) -b\Delta P(\theta) + c \eta(\theta),
\end{aligned}
\end{equation}
where $P(\theta)$, $\Delta P(\theta)$, and $\eta(\theta)$ represent the average power, fluctuations and efficiency of a cycle parameterized by a set of parameters $\theta$, and $\Sigma(P,\eta)$ is given by Eq.~(\ref{eq:sigma_p_eta}).

We wish to prove the following. 
Given three fixed scalars $a_1,b_1,c_1 > 0$, that do not necessarily sum to 1, let $\theta_1$ be the value of $\theta$ that locally maximizes $G(a_1,b_1,c_1;\theta)$. Then, it is always possible to identify three positive scalars $a_2,b_2,c_2>0$, such that the same parameters $\theta_1$ (i.e. the same cycle) is a local maximum for $F(a_2,b_2,c_2;\theta)$. In the following, we will use that
\begin{align}
	\partial_P \Sigma &\geq 0, & \partial_\eta \Sigma &< 0,
	\label{eq:s_cond}
\end{align}
and that the Hessian $H^{(\Sigma)}$ of $\Sigma(P,\eta)$ is given by
\begin{equation}
	H^{(\Sigma)} = 
	\begin{pmatrix}
		0 & -\beta_\text{C}\frac{\eta_\text{c}}{\eta^2} \\
		-\beta_\text{C}\frac{\eta_\text{c}}{\eta^2} & 2\beta_\text{C}P\frac{\eta_\text{c}}{\eta^3}
	\end{pmatrix}.
	\label{eq:s_hess}
\end{equation}

Proof: by assumption, $\theta_1$ is a local maximum for $G(a_1,b_1,c_1;\theta)$. Denoting with $\partial_i$ the partial derivative in $(\theta)_i$, we thus have
\begin{equation}
	0 = \partial_i G(a_1,b_1,c_1;\theta_1) = a_1 \partial_i P(\theta_1) - b_1 \partial_i\Delta P(\theta) + c_1 \partial_i \eta(\theta_1).
	\label{eq:partial_g}
\end{equation}
Now we compute the derivative in $\theta$ of $F(a_2,b_2,c_2;\theta)$, where $a_2,b_2,c_2>0$ are three arbitrary scalars, and we evaluate it in $\theta_1$. We have
\begin{equation}
	\partial_i F(a_2,b_2,c_2;\theta_1) = ( a_2 - c_2 \partial_P \Sigma )\partial_i P(\theta_1) -b_2\partial_i\Delta P(\theta) - (c_2\partial_\eta \Sigma)\partial_i \eta(\theta_1) .
\end{equation}
Therefore, if we choose $a_2,b_2,c_2$ such that 
\begin{equation}
\begin{pmatrix}
	a_1\\b_1\\c_1
\end{pmatrix}
=
\begin{pmatrix}
	1 & 0 & -\partial_P \Sigma \\
	0 & 1 & 0 \\
	0 & 0 & -\partial_\eta \Sigma
\end{pmatrix}
\begin{pmatrix}
	a_2\\b_2 \\ c_2
\end{pmatrix},
\label{eq:a_transf}
\end{equation}
thanks to Eq.~(\ref{eq:partial_g}) we have that 
\begin{equation}
	0 = \partial_i F(a_2,b_2,c_2;\theta_1),
\end{equation}
meaning that the same parameters $\theta_1$ that nullifies the gradient of $G$, nullifies also the gradient of $F$ at a different choice of the weights, given by Eq.~(\ref{eq:a_transf}).
The invertibility of Eq.~(\ref{eq:a_transf}) (i.e. a non-null determinant of the matrix) is guaranteed by Eq.~(\ref{eq:s_cond}). We also have to make sure that if $a_1,b_1,c_1>0$, then also $a_2,b_2,c_2>0$. To do this, we invert Eq.~(\ref{eq:a_transf}), finding
\begin{equation}
\begin{pmatrix}
	a_2\\b_2 \\c_2
\end{pmatrix}
=
\begin{pmatrix}
	1 & 0 & -\partial_P \Sigma/(\partial_\eta \Sigma) \\
	0& 1 & 0 \\
	0 & 0 &   -1/(\partial_\eta \Sigma)
\end{pmatrix}
\begin{pmatrix}
	a_1\\b_1\\c_1
\end{pmatrix}.
\label{eq:a_inv}
\end{equation}
It is now easy to see that also the weights $a_2,b_2,c_2$ are positive using Eq.~(\ref{eq:s_cond}).

To conclude the proof, we show that $\theta_1$ is a local maximum for $F(a_2,b_2,c_2;\theta)$ by showing that its Hessian is negative semi-definite. Since, by hypothesis, $\theta_1$ is a local maximum for $G(a_1,b_1,c_1;\theta)$, we have that the Hessian matrix
\begin{equation}
	H^{(G)}_{ij} \equiv \partial_{ij} G(a_1,b_1,c_1;\theta_1) = a_1 \partial_{ij}P -b_1\partial_{ij}\Delta P + c_1 \partial_{ij}\eta  
\end{equation}
is negative semi-definite. We now compute the Hessian $H^{(F)}$ of $F(a_2,b_2,c_2;\theta)$ in $\theta=\theta_1$:
\begin{equation}
	H^{(F)}_{ij} = a_2 \partial_{ij}P -b_2\partial_{ij}\Delta P - c_2\left[ \partial_P \Sigma\,\partial_{ij}P + \partial_\eta \Sigma\,\partial_{ij}\eta + Q_{ij}  \right],
	\label{eq:hg_def}
\end{equation}
where
\begin{equation}
	Q_{ij} =
	\begin{pmatrix}
		\partial_i P & \partial_i \eta
	\end{pmatrix}
	H^{(\Sigma)}
	\begin{pmatrix}
		\partial_j P \\ \partial_j \eta
	\end{pmatrix},
	\label{eq:qij_def}
\end{equation}
and	$H^{(\Sigma)}$ is the Hessian of $\Sigma(P,\eta)$ computed in $P(\theta_1)$ and $\eta(\theta_1)$.
Since we are interested in studying the Hessian of $F(a_2,b_2,c_2;\theta_1)$ in the special point $(a_2,b_2,c_2)$ previously identified, we substitute Eq.~(\ref{eq:a_inv}) into Eq.~(\ref{eq:hg_def}), yielding
\begin{equation}
	H^{(F)}_{ij} = H^{(G)}_{ij} + \frac{b_1}{\partial_\eta \Sigma} Q_{ij}.
\end{equation}
We now prove that $H^{(F)}_{ij}$ is negative semi-definite since it is the sum of negative semi-definite matrices. By hypothesis $H^{(G)}_{ij}$ is negative semi-definite. Recalling Eq.~(\ref{eq:s_cond}) and that $b_1 >0$, we now need to show that $Q_{ij}$ is positive semi-definite. Plugging Eq.~(\ref{eq:s_hess}) into Eq.~(\ref{eq:qij_def}) yields
\begin{equation}
	Q_{ij} = \beta_\text{C}\frac{\eta_\text{c}}{\eta^2}\partial_i\eta\,\partial_j\eta\,R_{ij},
\end{equation}
where
\begin{align}
	R_{ij} &\equiv 2P + S_{ij} + S^T_{ij}, &
	S_{ij} &= -\frac{\partial_i P}{\partial_i \eta}.
	\label{eq:r_def}
\end{align}
We now show that if $R_{ij}$ is positive semi-definite, then also  $Q_{ij}$ is positive semi-definite. By definition, $Q_{ij}$ is positive semi-definite if, for any set of coefficient $a_i$, we have that $\sum_{ij} a_i Q_{ij} a_j\geq 0$. Assuming $R_{ij}$ to be positive semi-definite, and using that $\beta_\text{C}, \eta_\text{c}, \eta >0$, we have 
\begin{equation}
	\sum_{ij} a_i Q_{ij} a_j = \beta_\text{C}\frac{\eta_
	\text{c}}{\eta^2} \sum_{ij} x_i R_{ij}x_j \geq 0,
\end{equation}
where we define $x_i \equiv \partial_i \eta \,a_i$. We thus have to prove the positivity of $R_{ij}$. We prove this showing that it is the sum of $3$ positive semi-definite matrices. Indeed, the first term in Eq.~(\ref{eq:r_def}), $2P$, is proportional to a matrix with $1$ in all entries. Trivially, this matrix has $1$ positive eigenvalue, and all other ones are null, so it is positive semi-definite. At last, $S_{ij}$ and its transpose have the same positivity, so we focus only on $S_{ij}$. $S_{ij}$ is a matrix with all equal columns. This means that it has all null eigenvalues, except for a single one that we denote with $\lambda$. Since the trace of a matrix is equal to the sum of the eigenvalues, we have $\lambda = \mathrm{Tr}[S]=\sum_i S_{ii}$. Using the optimality condition in Eq.~(\ref{eq:partial_g}), we see that each entry of $S$ is positive, i.e. $S_{ij}>0$. Therefore $\lambda >0$, thus $S$ is positive semi-definite, concluding the proof that $H^{(F)}_{ij}$ is negative semi-definite.

To conclude, we notice that we can always renormalize $a_2,b_2,c_2$, such that they sum to $1$, preserving the same exact optimization problem.

\subsection{Identifying optimal cycles with reinforcement learning}
\label{sec:rl}
In this subsection we show how the formulation of the optimization problem given in Sec.~\ref{sec:extended_state} in terms of an extended state allows us to use RL to optimize a figure of merit containing power fluctuations, and we provide details on the RL implementation.
The RL method that we use is based on the Soft Actor-Critic algorithm \cite{haarnoja2018_pmlr}, introduced in the context of robotics and video-games \cite{haarnoja2018_arxiv_sac, haarnoja2018_arxiv_walk, christodoulou2019, delalleau2019}, generalized to optimize multiple objectives.
RL has received great attention for its success at mastering tasks beyond human-level such as playing games \cite{mnih2015,silver2017,vinyals2019}, for robotic applications \cite{haarnoja2018_arxiv_sac, haarnoja2018_arxiv_walk}, and for controlling plasmas in a tokamak \cite{degrave2022}. RL has been recently used for quantum control \cite{bukov2018,an2019,dalgaard2020,mackeprang2020,schafer2020,schafer2021,porotti2021,metz2022}, outperforming previous state-of-the-art methods \cite{niu2019,zhang2019}, for fault-tolerant quantum computation \cite{fosel2018,sweke2020}, and in the field of quantum thermodynamics \cite{ ashida2021,erdman2022,sgroi2021,erdman2022_arxiv}.  RL also allows to optimize ``blackbox systems'', i.e. to perform an optimization without requiring any explicit knowledge of the dynamics of the system being optimized, thus potentially applicable to experimental devices \cite{sgroi2021,erdman2022_arxiv}. In light of this, we expect RL to be a powerful tool also in this context. Furthermore, since RL is based on the Markovian Decision Process framework (see below and Ref.~\cite{sutton2018} for details), it is a natural choice for systems described by Markovian dynamics.
 In principle, also other optimization methods, such as Pontryagin’s Minimum Principle, could be used \cite{kirk2004}. However, this method only provides, in general, necessary conditions for an optimal control. Furthermore, it could in practice get stuck in local maxima, and it requires some hand-tuning in the case of non-analytic controls. As we see from our results, optimal cycles are often given by of piece-wise continuous function, which are not analytic. This would complicate the use of Pontryagin’s Minimum Principle.

The method here described is an extension of the methods described in Refs.~\cite{erdman2022, erdman2022_arxiv}. 
We therefore refer to the Ref.~\cite{erdman2022} for an explanation of the RL method, and we adopt its notation to explain all the differences and generalizations that we put forward in this paper.

First, we define the optimization problem detailing the choice of the \textit{state space}, the \textit{action space}, and of the \textit{reward}. Let us discretize time in time steps $t_i = i\Delta t$. In Ref.~\cite{erdman2022} just the power is optimized, so the state of the environment is described by the density matrix $\rho$. Indeed, RL is based on the Markov Decision Process assumption, meaning that the \textit{reward} at a given time-step must be a function of the last state and action. In Ref.~\cite{erdman2022}, the \textit{reward} is the power averaged over the last time-step, which can be computed from the density matrix and from the value of the control at the last time-step. However, power fluctuations cannot be computed solely from the knowledge of $\rho$.
Here, to optimize also power fluctuations, we employ the extended state as state of the environment, i.e. we choose as \textit{state space} $\mathcal{S} = \{(\rho,s,\vec{u})| \rho\in\mathcal{D}, s\in \mathcal{E}, \vec{u}\in \mathcal{U}  \}$, where $\mathcal{D}=\{ \rho| \rho \geq 0, \Tr[\rho] =1 \}$ is the space of density matrices, $\mathcal{E}=\{s|s=s^\dagger, \Tr[s]=0\}$ is the space of traceless Hermitian operators, and $\mathcal{U}$ is the continuous set of allowed control parameters. The state at time-step $i$ is then given by $s_{i}=(\rho_{t_{i}}, s_{t_{i}}, \vec{u}(t_{i-1}))$. The \textit{action space} is given by $\mathcal{A}=\{(\vec{u},\beta)|\vec{u}\in\mathcal{U}, \beta\in[\beta_\text{H},\beta_\text{C}]  \}$, such that the action at time-step $t_i$ is $a_i=(\vec{u}_i,\beta_i)$. The controls $\vec{u}(t)$ and $\beta(t)$ are then kept constant at $\vec{u}_i$ $\beta_i$ for the current time-step. The \textit{reward} is given by
\begin{equation}
    r_{i} = \frac{1}{\Delta t} \int_{t_{i-1}}^{t_i}\left[ a\frac{{P}(t)}{P_\text{max}} -b\frac{\Delta{P}(t)}{\Delta P(P_\text{max})} - c\frac{\Sigma(t)}{\Sigma(P_\text{max})}\right]dt,
\end{equation}
which corresponds to the figure of merit averaged only over the last time-step. Together with Eqs.~ (\ref{eq:p_sigma}) and (\ref{eq:dp}) of the main text, we see that this choice satisfies the Markov Decision Process assumption.

As in Ref.~\cite{erdman2022}, we formulate the optimization problem as a discounted, continuing RL task, where the aim is to learn a policy $\pi(a|s)$ that, at each time step $t_i$, maximizes the \textit{return}, i.e. the long-term average of the rewards:
\begin{equation}
    r_{i+1} + \gamma r_{i+2} + \gamma^2 r_{i+3} + \dots = \sum_{k=0}^\infty \gamma^k r_{i+1+k},
\end{equation}
where $\gamma \in [0,1)$ is the discount factor which determines how much we are interested in future rewards, as opposed to immediate rewards. By choosing $\gamma$ close enough to $1$, we are optimizing the figure of merit averaged over a long time-scale, such that the method should automatically discover to perform cycles (see Refs.~\cite{erdman2022,erdman2022_arxiv} for additional details).

In this work, we employ the soft actor-critic method \cite{haarnoja2018_arxiv_sac,haarnoja2018_arxiv_walk} as implemented in Ref.~\cite{erdman2022} with the following changes.
\begin{itemize}
    \item \textbf{Policy parameterization}. Here we do not have a discrete action, but we have multiple continuous actions given by $\vec{\xi} \equiv (\vec{u},\beta)$. We therefore need to parameterize the policy $\pi(\vec{\xi}| s)$ as a multi-variate distribution function. We use the same approach as in Ref.~\cite{erdman2022}, but we replace the normal distribution with a multivariate normal distribution.
    To this end, we employ a Neural Network (NN) that takes the state $s$ as input, and outputs a vector $\vec{\mu}$ and a matrix $M$. We use a multilayer perceptron NN with $2$ hidden layer. We then produce samples from the multivariate normal distribution with mean $\vec{\mu}$ and covariance matrix $M^TM + \lambda\mathbb{1}$, where $\lambda > 0$ is a small real number added for numerical stability. Then, as in Ref.~\cite{erdman2022}, we determine a sample of $\pi(\vec{\xi}|s)$ applying a ``squash function'', in this case a hyperbolic tangent, to each variable of the multivariate distribution. This ensures that each action lies in the specified continuous interval $\mathcal{U}$.

    \item \textbf{Automatic ``temperature'' tuning.} In the soft-actor critic method, the exploration-exploitation balance is governed by the hyper-parameter $\varepsilon$, known in the RL literature as the ``temperature'' parameter. In Ref.~\cite{erdman2022}, $\varepsilon$ was scheduled during training. However, its value depends on the magnitude of the reward, making it quite model-dependent. Here, instead, we follow Ref. \cite{haarnoja2018_arxiv_sac} to automatically tune $\varepsilon$ during training. The idea is to change $\varepsilon$ during training as to set the average entropy of the policy to a target value $\bar{H}$. Large values of $\bar{H}$ produce an exploratory policy, while smaller values of $\bar{H}$ produce a more deterministic policy. As detailed in Ref.~\cite{haarnoja2018_arxiv_sac} and implemented in Ref.~\cite{erdman2022_arxiv}, the tuning of $\varepsilon$ is done by minimizing the following loss function
    \begin{equation}
        L_H(\varepsilon) \equiv \varepsilon \underset{ \substack{s\sim \mathcal{B} }}{\text{E}}\left[  H(\pi(\cdot|s)) - \bar{H} \right],
    \end{equation}
    where $H(P) \equiv \underset{ \substack{x\sim P }}{\text{E}}[-\ln P(x)]$ is the entropy of the distribution $P(x)$. For numerical stability, if $\varepsilon$ becomes negative during training, we reset it to a small positive number. To favour exploration in the early phases of the training, while still obtaining a final policy that is nearly deterministic, we exponentially schedule the target entropy $\bar{H}$ during training as follows
    \begin{equation}
        \bar{H}(n_\text{steps}) = \bar{H}_\text{end} + (\bar{H}_\text{start}-\bar{H}_\text{end}) \exp(-n_\text{steps}/\bar{H}_\text{decay}),
    \end{equation}
    where $\bar{H}_\text{start}$, $\bar{H}_\text{end}$ and $\bar{H}_\text{decay}$ are training hyperparameters.
    
    \item
    \textbf{Training steps}. The training steps are performed as in Ref.~\cite{erdman2022}, i.e. repeatedly using the ADAM algorithm to minimize the loss functions $L_\pi$ and $L_Q$ computed over a batch of experience drawn from the replay buffer $\mathcal{B}$. Here, in addition, every time an optimization step of $L_\pi$ and $L_Q$ is performed, we also update $\varepsilon$ performing one optimization step of $L_H$ computed over the same batch of experience.
    
    \item
    \textbf{Optimal cycle evaluation.} Once the training is complete, all cycles and values of the power, power fluctuations, entropy production and efficiency reported in the main text are computed evaluating the deterministic policy. More specifically, we determine the optimal cycle using the deterministic policy, i.e. choosing actions according to the \textit{mean} of the multivariate normal distribution, instead of sampling from it. This turns the stochastic policy into a deterministic one that typically performs better.

\end{itemize}

\section{Fast-Driving for generic systems}
\label{app:Fast_Driving}

In this section we derive analytic expressions for the main quantities~\eqref{eqapp:P_dP_Sigma} in the regime of fast-driving, i.e. when the period $\tau$ of the cycle is small compared to the relaxation timescales of the system. As this regime is rigorously justified for stochastic dynamics, and for easiness of presentation, we will assume diagonal operators that commute at all times (cf. Ref.~\cite{cavina2021}). 

First, for what concerns power and entropy production, notice, from~\eqref{eqapp:P_dP_Sigma} that both can be computed from the integral of
\begin{align}
    \Tr[\dot{\rho}_t H_{\vec{u}(t)}]=\Tr[\mathcal{L}_{\vec{u}(t),\beta(t)}[\rho_t] H_{\vec{u}(t)}]\;.
\end{align}
In the fast driving approximation, the cycle is much quicker than the relaxation time of the system. The consequence is that $\rho_t\simeq \rho^{(0)}$ tends to a steady-state solution, which can be computed as the leading order of a perturbative expansion~\cite{cavina2021}. It follows that power and entropy production can be computed as
\begin{align}
     \ev{P} &=\frac{1}{\tau} \int_0^\tau dt \Tr[\mathcal{L}_{\vec{u}(t),\beta(t)}[\rho^{(0)}] H_{\vec{u}(t)}], \\
    \ev{\Sigma} &= -\frac{1}{\tau} \int_0^\tau dt \Tr[\mathcal{L}_{\vec{u}(t),\beta(t)}[\rho^{(0)}] H_{\vec{u}(t)}]\beta(t).
\end{align}
The generic expression for $\rho^{(0)}$ is given in Ref.~\cite{cavina2021}. At the same time, for the purposes of the present work, it is enough to consider the simple case in which the dynamics is described by a scalar master equation $\mathcal{L}[...] = \gamma\pi_t\Tr[...]-\gamma\mathbb{1}$, where
\begin{align}
    \pi_t=\frac{e^{-\beta(t) H_{\vec{u}(t)}}}{\Tr[e^{-\beta(t) H_{\vec{u}(t)}}]}
\end{align}
is the instantaneous steady state and $\gamma^{-1}$ the relaxation timescale. In such case it is clear that, as $\rho^{(0)}$ is the steady state, it satisfies
\begin{align}
    \int_0^\tau dt \gamma\left(\pi_t -\rho^{(0)}\right)=0\;,
\end{align}
leading to
\begin{align}
    \rho^{(0)}=\frac{\int_0^\tau dt\ \pi_t}{\tau}=\ev{\pi}\;,
\end{align}
and therefore
\begin{align}
\label{eqapp:P_fast}
     \ev{P} &=\frac{1}{\tau} \int_0^\tau dt \gamma \Tr[(\pi_t-\ev{\pi}) H_{\vec{u}(t)}], \\
\label{eqapp:Sigma_fast}
    \ev{\Sigma} &= -\frac{1}{\tau} \int_0^\tau dt \gamma\Tr[(\pi_t-\ev{\pi}) H_{\vec{u}(t)}]\beta(t).
\end{align}

The most general expression for the heat current in the fast-driving regime is given in Ref.~\cite{cavina2021}. In the following, we derive new analytical expressions that allow to efficiently compute also the power fluctuations in the same regime.

\subsection{Fluctuations in the Fast driving regime for arbitrary stochastic (classical) systems}
As mentioned above, we will simplify the discussion by considering semi-classical systems represented by commuting operators at all time.

\subsubsection{Simple Lindbladian}
First, we start again by considering the simple scalar master equation $\mathcal{L}[...] = \gamma\pi_t\Tr[...]-\gamma\mathbb{1}$, with $
    \pi_t=e^{-\beta(t) H_{\vec{u}(t)}} / \Tr[e^{-\beta(t) H_{\vec{u}(t)}}]
$. 
In such case, the equations for $\rho_t$~\eqref{eq:lindblad} and $s_t$~\eqref{eq:s_ode} become (from now on we simplify the notation as $H_{\vec{u}(t)}\equiv H_t$)
\begin{align}
    \dot{\rho}_t=& \gamma(\pi_t-\rho_t)\;,\label{eq:master_eq_sm}\\
    \dot{s}_t=& -\gamma s_t+2\rho_t\dot{H}_t-2\Tr[\rho_t\dot{H}_t]\rho_t \equiv -\gamma s_t + 2\rho_t{\tilde{\dot{H}}_\rho}_t\;,
\end{align}
where we defined 
\begin{align}
\label{eqapp:tilde_def}
    \tilde{A}_\rho= A-\rho\Tr[\rho A]\;.
\end{align}
The asymptotic solutions for $\pi_t$ and $s_t$, in the case of periodic driving with period $\tau$, are
\begin{align}
\label{eqapp:rho_simp_exact}
    \rho_t=& \frac{\int_{t-\tau}^t dt'\; \gamma \pi_{t'} e^{\gamma(t'-t)}}{1-e^{-\gamma\tau}}\;,
    \\
\label{eqapp:s_simp_exact}
    s_t=& \frac{\int_{t-\tau}^t dt'\; 2\rho_{t'}{\tilde{\dot{H}}_\rho}_{t'} e^{\gamma(t'-t)}}{1-e^{-\gamma\tau}}\;.
\end{align}
Notice that until now, no approximation has been used. To compute all relevant quantities, we will use the fast-driving condition $\gamma\tau\ll 1$ and expand in powers of $\gamma\tau$. That is,
\begin{align}
    \rho_t=\rho^{(0)}_t+\rho^{(1)}_t+\dots \quad \rho^{(i)}_t\sim \mathcal{O}(\gamma^i\tau^i)\;,\\
    s_t=s^{(0)}_t+s^{(1)}_t+\dots \quad s^{(i)}_t\sim \mathcal{O}(\gamma^i\tau^i)\;.
\end{align}
It is easy to verify, from~\eqref{eqapp:rho_simp_exact},
\begin{align}
    \rho^{(0)}_t &\equiv\rho^{(0)}=\ev{\pi}\;,\\
    \dot{\rho}^{(0)}_t &=0\;,\\
    \dot{\rho}^{(1)}_t &=\gamma(\pi_t-\rho^{(0)})\;.
\end{align}
The expansion of $s_t$ is slightly more complex, but for our purposes we only need $s^{(0)}$, which can be found integrating by parts~\eqref{eqapp:s_simp_exact}, 
\begin{align}
     s_t= \frac{\int_{t-\tau}^t ds\; 2\rho_s {\tilde{\dot{H}}_\rho}_s e^{\gamma(s-t)}}{1-e^{-\gamma\tau}}
     = 2\rho_t 
     {{\tilde{H}}_{\rho_t}} -
     \frac{2}{1-e^{-\gamma\tau}}\int_{t-\tau}^t dt'\left( \dot{\rho}_{t'}{\tilde{H}_{\rho_{t'}}} e^{\gamma(t'-t)}+
     \gamma \rho_{t'}{\tilde{H}_{\rho_{t'}}} e^{\gamma(t'-t)}- \rho_{t'}\Tr[H_{t'} \dot{\rho}_{t'}]e^{\gamma(t'-t)}\right) \;.
\end{align}
In this expression we can substitute all the quantities at the leading order, obtaining, using the fluctuations notation $\delta A:=A-\langle A\rangle$, 
\begin{align}
    s^{(0)}_t=2\rho^{(0)}\tilde{H}_{\rho^{(0)}}-\frac{2}{\gamma\tau}\int_{t-\tau}^t dt' \left( \gamma\delta\pi_{t'}\tilde{H}_{\rho^{(0)}}+\gamma\rho^{(0)}\tilde{H}_{\rho^{(0)}}-\gamma\rho^{(0)}\Tr[H_{t'}\delta\pi_{t'}]\right)
\end{align}
Notice that although $\rho^{(0)}$ is constant, the quantity $\tilde{H}_{\rho^{(0)}}$ (cf. Eq.~\eqref{eqapp:tilde_def}) is time-dependent as $H_t$ is time dependent.
The above expression can be further simplified by dropping all the time-dependence in the notation and expanding the time-average at leading order
\begin{align}
\label{eqapp:s0}
    s^{(0)}=2\left(
    \rho^{(0)} \tilde{H}_{\rho^{(0)}}- \langle \rho^{(0)} \tilde{H}_{\rho^{(0)}}\rangle 
    \right)+2\bigg\langle \rho^{(0)}\Tr[\delta H \delta \pi] +\delta\pi \Tr[\delta H \rho^{(0)}] - \delta\pi \delta H \bigg\rangle\;.
\end{align}

\paragraph*{\bf{Fluctuations.}} The fluctuations become, integrating by parts~\eqref{eqapp:P_dP_Sigma},
\begin{align}
\label{eq:fluctt}
    \Delta P=\frac{1}{\tau}\int_0^\tau dt \Tr[\dot{H}_t s_t]=-\frac{1}{\tau}\int_0^\tau dt \Tr[H_t\dot{s}_t] =\frac{1}{\tau}\int_0^\tau dt \Tr[H_t(\gamma s_t-2\tilde{\dot{H}}_{\rho_t})]\;.
\end{align}
Therefore we want to compute
\begin{align}
    I_2=& -\frac{1}{\tau}  \int_0^\tau dt \Tr[2H_t\tilde{\dot{H}}_{\rho_t}]\;,\\
    I_1=& \frac{1}{\tau}\int_0^\tau dt \Tr[\gamma H_t s_t]\;,
\end{align}
at leading order.
Let's compute the two terms
\begin{multline}
    I_2= -\frac{1}{\tau}\int_0^\tau dt \Tr[2H_t\tilde{\dot{H}}_{\rho_t}]=-\frac{1}{\tau}\int_0^\tau dt \Tr[2H_t(\dot{H}_t\rho_t-\rho_t\Tr[\dot{H}_t\rho_t])]=
    \frac{1}{\tau}\int_0^\tau dt  \Tr[H_t^2\dot{\rho}_t-2\Tr[\rho_t H_t]\Tr[\dot{\rho}_t H_t]]=\\
    =\gamma\left( 
    \langle \Tr[H^2\delta\pi]\rangle -2\langle \Tr[H\delta\pi]\Tr[H \rho^{(0)}]\rangle
    \right)\;,
\end{multline}
where for the last equality we used the equation of motions $\dot{\rho}_t=\gamma(\pi_t-\rho_t)$ a the leading order, where $\rho_t\sim\rho^{(0)}=\langle\pi\rangle$. We remind that we use the notation $\delta A:=A-\langle A\rangle$.
Notice that for $I_1$ we only need $s$ at order $\mathcal{O}(\gamma^0\tau^0)$.
$I_1$ can therefore be computed as
\begin{align}
    I_1=\gamma \langle \Tr[H s^{(0)}]\rangle\;,
\end{align}
and the total fluctuations are, after substituting $s^{(0)}$~\eqref{eqapp:s0} and some tedious algebra,
\begin{multline}
\label{eq:DeltaP_cov}
    \frac{\Delta P}{2\gamma}= \frac{I_2 + I_1}{2\gamma}=\\
    =\ev{ \Tr[\rho^{(0)} H^2] - \Tr[\rho^{(0)} H]^2 } - \left( \Tr[\rho^{(0)}\ev{H}^2]  - \Tr[ \rho^{(0)}\ev{H} ]^2 \right)
    +\frac{1}{2}\langle \Tr[\delta H^2\delta \pi]\rangle -\langle \Tr[\delta H \delta\pi]\Tr[\delta H \rho^{(0)}]\rangle \;.
\end{multline}
Under the same assumptions, the power~\eqref{eqapp:P_fast} can be expressed as
\begin{equation}
    P = \gamma \langle \Tr[(\pi-\langle\pi\rangle)(H-\langle H\rangle)] \rangle \equiv \gamma \Tr[ \text{Cov}[\pi, H] ],
    \label{eq:p_cov}
\end{equation}
and similarly the entropy production~\eqref{eqapp:Sigma_fast}
\begin{equation}
    \Sigma = \gamma \langle \Tr[(\pi-\langle\pi\rangle)(\beta H-\langle \beta H\rangle)] \rangle \equiv \gamma \Tr[ \text{Cov}[\pi,\beta H] ],
    \label{eq:p_cov2}
\end{equation}
where the covariance is with respect to time.

\subsubsection{General Stochastic Classical Case}
Moving to the case of a general Lindbladian, the expression for power and entropy production can be found as from Reference~\cite{cavina2021}. For what concerns the fluctuations, these can be again expressed as (see~\eqref{eq:fluctt}) 
\begin{align}
    \Delta P=-\frac{1}{\tau}\int_0^\tau dt \Tr[H_t\dot{s}_t] =-\frac{1}{\tau}\int_0^\tau dt \Tr[H_t(\mathcal{L}_t[s_t] +2\tilde{\dot{H}}_{\rho_t})]=I_1+I_2\;.
\end{align}
In the next expressions we drop all the time-dependencies to simplify the notation.
The computation of $I_1$ and $I_2$ is in this case
\begin{align}
    I_2=& -\frac{1}{\tau}\int_0^\tau dt \Tr[2H\tilde{\dot{H}}_\rho]=\langle \Tr[H^2\mathcal{L}[\rho^{(0)}]]\rangle - 2\langle \Tr[H\mathcal{L}[\rho^{(0)}]]\Tr[H\rho^{(0)}]\rangle\;,\\
    I_1=& -\frac{1}{\tau}\int_0^\tau dt \Tr[H \mathcal{L}[s_0]]\;,
\end{align}
where the first expression can be obtained integrating by parts, and everything is expanded at the leading order in $|\mathcal{L}|\tau$, $\rho^{(0)}$ is the steady state of the fast driving (see \cite{cavina2021} for an analytical expression) and $s_0$ is the leading order term of $s$, which can be computed from the analytical solution
\begin{align}
    s_t=\int_{-\infty}^t dt'\;  U_{t,t'}[ 2\rho_{t'}\tilde{\dot{H}}_{\rho_{t'}}]\;,
\end{align}
$U_{t,t'}$ being the propagator, i.e. time-ordered exponential of $\mathcal{L}$ between $t'$ and $t\geq t'$. Expanding the leading term one obtains
\begin{align}
    s_0=2\rho^{(0)}\tilde{H}_{\rho^{(0)}}+2\langle\mathcal{L}\rangle^{-1}
    \left( \langle \mathcal{L}[\rho^{(0)}]\tilde{H}_{\rho^{(0)}}\rangle - \rho^{(0)} \Tr[\langle H\mathcal{L}[\rho^{(0)}]\rangle]
    -\langle\mathcal{L}[\rho^{(0)}\tilde{H}_{\rho^{(0)}}]\rangle\right)\;.
\end{align}

\section{Quantum Dot Based Heat Engine}

\subsection{Optimality of the Otto cycle in the fast driving for power and entropy production trade-offs}
Here we prove that, among all possible cycles described by $\beta(t)$ and $u(t)$, Otto cycles in the fast driving regime maximize an arbitrary trade-off between power and entropy production. 
The proof stems from a simple generalization of Appendix A of Ref.~\cite{erdman2019_njp}, where it is shown that, among all possible Otto cycles, Otto cycles in the fast driving regime maximize the power.

To this end, we consider the figure of merit $\ev{F}$ with $b=0$:
\begin{equation}
    \ev{F} = a\frac{\ev{P}}{P_\text{max}} -c \frac{\ev{{\Sigma}}}{\Sigma(P_\text{max})}.
\end{equation}
By neglecting fluctuations, we no not need the extended state, so we can write $\ev{F}$ as the time average of a function of the state $p_t$ (defined in the main text) and of the controls $\beta(t)$ and $u(t)$, i.e. as
\begin{equation}
    \ev{F} =\frac{1}{\tau}\int_0^\tau f(p_t, u(t),\beta(t)) dt,
\end{equation}
where $f(p,u,\beta)$ is a suitable function.
This assumption is sufficient to use the argument of \cite{erdman2019_njp}, with $\ev{F}$ replacing the average power,  to prove that optimal cycles are Otto cycles in the fast-driving regime. Here we outline the general idea, referring to \cite{erdman2019_njp} for additional details.
As argued here and in \cite{erdman2019_njp}, given a cycle with period $\tau$, also the state $p_t$ will become periodic with the same period $\tau$. We can thus represent an arbitrary cycle cycle as a closed curve in the $u-p$ plane and in the $\beta-p$ plane. It can be shown that, given any fixed cycle $u(t)$ and $\beta(t)$, we can always define two sub-cycles such that one of the two has a larger or equal $\ev{F}$ than the original one. The two sub-cycles are defined by introducing quenches in the control, i.e. ``cutting vertically'' the cycle represented in the $u-p$ and $\beta-p$ plane. By reiterating this process over and over, we end up with an Otto cycle in the fast driving regime, thus concluding the proof.

\subsection{Fast driving regime}

\begin{figure}[!tb]
	\centering
	\includegraphics[width=0.5\columnwidth]{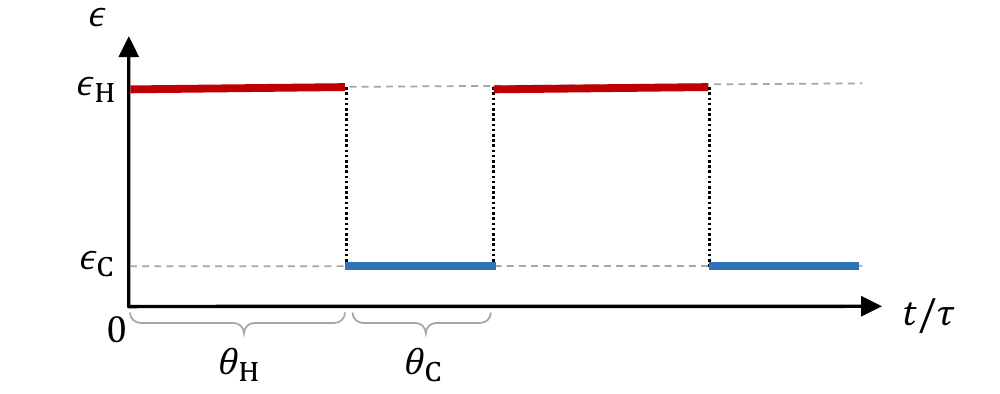}
	\caption{Schematic representation of the Otto cycle. The value of the gap $\epsilon(t) \equiv u(t)E_0$ is plotted as a function of $t/\tau$. The red segment corresponds to setting $\beta(t)=\beta_\text{H}$, and the blue segment to $\beta(t) = \beta_\text{C}$.}
	\label{fig:otto}
\end{figure}

Using Eqs.~(\ref{eq:DeltaP_cov},\ref{eq:p_cov},\ref{eq:p_cov2}), we can easily compute the power, entropy production and fluctuations of a generic Otto cycle in the fast driving regime. We fix an Otto cycle as in Fig.~\ref{fig:otto}, where $\epsilon_\text{H}$ and $\epsilon_\text{C}$ are the values of $u(t)E_0$ while in contact respectively with the hot (H) and cold (C) bath, and $\theta_\text{H}$ and $\theta_\text{C}$ represent the time fraction of the cycle spent in contact with the respective bath ($\theta_\text{C}+\theta_\text{H}=1$).
This yields
\begin{equation}
\begin{aligned}
    \ev*{P} &= \gamma   \theta_\text{H}\theta_\text{C}\, (\epsilon_\text{H}-\epsilon_\text{C})(f_\text{H}-f_\text{C}), \\
    \ev*{\Delta P} &= 2\gamma\,\theta_\text{H}\theta_\text{C}\, (\epsilon_\text{H}-\epsilon_\text{C})^2 \left[ \bar{f} (1-\bar{f}) + (\frac{f_\text{C}+f_\text{H}}{2} -\bar{f})(1-2\bar{f}) \right], \\
    \ev*{\Sigma} &= -\gamma\theta_\text{H}\theta_\text{C}(\beta_\text{H}\epsilon_\text{H}-\beta_\text{C}\epsilon_\text{C})(f_\text{H}-f_\text{C}),
\end{aligned}
\label{eq:qd_otto}
\end{equation}
where we define $f_\alpha = f(\beta_\alpha\epsilon_\alpha)$, for $\alpha=\text{H},\text{C}$ and the average excited state occupation
$ \bar{f} = \theta_\text{H}f_\text{H} + \theta_\text{C}f_\text{C}$.
 Without loss of generality, in the heat engine regime we can assume $\epsilon_\alpha \geq 0$. This condition, together with the heat engine condition $P\geq 0$, can be expressed as 
\begin{equation}
    0 \leq x_\text{H} \leq x_\text{C} \leq x_\text{H}(1+dT),
    \label{eq:app_engine_condition}
\end{equation}
where we introduce the dimensionless quantities $x_\alpha = \beta_\alpha\epsilon_\alpha$, and where we define the dimensionless temperature difference
\begin{equation}
    dT \equiv \beta_\text{C}(\beta_\text{H}^{-1}-\beta_\text{C}^{-1}).
\end{equation}
For simplicity, in this appendix we define a figure of merit $G$ with a different normalization, i.e.
\begin{equation}
    \ev*{G} = a\left(\frac{\ev*{P}}{\gamma T}\right) -b\left(\frac{\ev*{\Delta P}}{\gamma T^2}\right) -c\left(\frac{\ev*{\Sigma}}{\gamma}\right),
    \label{eq:g_figmerit_def}
\end{equation}
where $T=\beta_\text{C}^{-1}$ and $k_B=1$.

We now assume that the temperature difference is small, i.e. that $dT\ll 1$. We therefore study $\ev*{G}$ to leading order in $dT$. We carry out the optimization with respect to the time fractions $\theta_\alpha$ and with respect to the dimensionless energy gaps $x_\alpha$.
We expand $x_\text{C}$ as 
\begin{equation}
    x_\text{C} = x_\text{H} (1+\delta x_\text{C}  ).
    \label{eq:xc_exp}
\end{equation}
 Imposing the heat engine condition in Eq.~(\ref{eq:app_engine_condition}), we find that $\delta x_\text{C}$  must satisfy 
 \begin{equation}
     0 \leq \delta x_\text{C} \leq dT;
     \label{eq:dxc_condition}
 \end{equation}
 therefore, $\delta x_\text{C}$ is a first order quantity in $dT$.
Expanding $\ev*{P}$, $\ev*{\Delta P}$ and $\ev*{\Sigma}$ to leading order in $dT$, we find
\begin{equation}
\begin{aligned}
	\ev*{P}/(\gamma T) &= \theta_\text{C}\theta_\text{H} g(x_\text{H})\delta x_\text{C}(dT-\delta x_\text{C}), \\
	\ev*{\Delta P}/(\gamma T^2) &= 2\theta_\text{C}\theta_\text{H} g(x_\text{H})(dT-\delta x_\text{C})^2, \\
	\ev*{\Sigma}/\gamma &= \theta_\text{C}\theta_\text{H} g(x_\text{H})\delta x_\text{C}^2,
\end{aligned}
\label{eq:p_dp_entropy_second}
\end{equation}
where we define
 \begin{equation}
     g(x) \equiv \frac{x^2}{2(1+\cosh{x})}.
     \label{eq:g_def}
 \end{equation}
Plugging this expansion into Eq.~(\ref{eq:g_figmerit_def}) yields
\begin{equation}
    \ev*{G} = \theta_\text{H}\theta_\text{C} g(x_\text{H}) \left[ \delta x_\text{C}\left( a dT - (a+c)\delta x_\text{C} \right) -2b\left( dT -\delta x_\text{C} \right)^2  \right].
    \label{eq:app_h1}
\end{equation}

First, we notice that $\ev*{P}$, $\ev*{\Delta P}$ and $\ev*{\Sigma}$ in Eq.~(\ref{eq:p_dp_entropy_second}), valid in the small temperature difference regime, exactly saturate the steady-state thermodynamic uncertainty relation even before performing any optimization, i.e. $\xi=1$, where, combining Eqs.~(\ref{eq:tur_ineq}) of the main text and (\ref{eq:eta})
\begin{equation}
    \xi = 2\frac{\ev*{P}^2}{\ev*{\Sigma}\ev*{\Delta P}}.
\end{equation}

We now maximize $\ev*{G}$ as written in Eq.~(\ref{eq:app_h1}). The optimization over $\theta_\alpha$ and $x_\text{H}$ depends on the term in the square parenthesis. We therefore distinguish two cases:

\subsubsection{Case 1}
If the term in square parenthesis is non-null, then we trivially have that $\theta_\alpha=1/2$ and $x_\text{H}=x_\text{max}$, where $x_\text{max}$ is the value that maximizes $g(x)$. It is implicitly defined by $x_\text{max}>0$ and by solving the transcendental equation
 \begin{equation}
     x_\text{max}\tanh(x_\text{max}/2) = 2.
     \label{eq:xh_ott}
 \end{equation}
We can then determine $\delta x_\text{C}$ by taking the derivative of $G$ and setting it to zero. This yields
\begin{equation}
    \delta x_\text{C}^{*1} = \frac{a+4b}{2a+4b+2c}dT.
\end{equation}
It is easy to see that this solution always satisfies the heat engine condition in Eq.~(\ref{eq:dxc_condition}). In this point we have that
\begin{equation}
    \ev*{G}^{*1} = g_\text{max}\frac{a^2-8bc}{16(a+2b+c)} dT^2,
\end{equation}
where we define $g_\text{max}\equiv g(x_\text{max}) \approx 2.40$, and
\begin{equation}
\begin{aligned}
	\ev*{P}^{*1}/(\gamma T) &= g_\text{max}\frac{(a+4b)(a+2c)}{16(a+2b+c)^2}dT^2, & ~ \quad ~
	\ev*{\Delta P}^{*1}/(\gamma T^2) &= g_\text{max}\frac{(a+2c)^2}{8(a+2b+c)^2}dT^2, \\
	\ev*{\Sigma}^{*1}/\gamma &= g_\text{max}\frac{(a+4b)^2}{16(a+2b+c)^2}dT^2, &
	\frac{\eta^{*1}}{\eta_\text{c}} &= \frac{a+2c}{2a+4b+2c}.
\end{aligned}
\label{eq:p_dp_entropy_1}
\end{equation}
We thus see that if $a^2-8bc>0$, then the figure of merit is positive. In the opposite case, the figure of merit is negative, so doing nothing becomes more convenient (since doing nothing, or doing Carnot cycles, gives $\ev*{G}=0$). Notice that if $a^2=8bc$, then any value of $\theta_\alpha$ and $x_\text{H}$ gives the same figure of merit, so all these solutions must be included in the Pareto front.

Setting $a=1$ and $b=c=0$, we find the maximum power solution:
\begin{align}
    \frac{P_\text{max}}{\gamma T} &= \frac{g_\text{max}}{16} dT^2, &
    \frac{\Delta P(P_\text{max})}{\gamma T^2} &= \frac{g_\text{max}}{8} d T^2, &   \frac{\Sigma(P_\text{max})}{\gamma} &= \frac{g_\text{max}}{16}dT^2,  &  \frac{\eta}{\eta_\text{c}} &= \frac{1}{2}.
    \label{eq:p_dp_max}
\end{align}
These relations imply
\begin{equation}
    \Delta P(P_\text{max}) = 2T P_\text{max}.
    \label{eq:max_rel}
\end{equation}

Setting $c=0$ and $b=1-a$ in Eq.~(\ref{eq:p_dp_entropy_1}), we can find the Pareto front between power and power fluctuations, i.e. the outer border of Fig.~\ref{fig:full_pareto} of the main text. After some algebra, it can be shown that
\begin{equation}
    \frac{\ev*{P}}{P_\text{max}} = 2\sqrt{\frac{\ev*{\Delta P}}{\Delta P(P_\text{max})}}- \frac{\ev*{\Delta P}}{\Delta P(P_\text{max})}.
    \label{eq:pareto_analytic}
\end{equation}
This corresponds to the black border shown in Fig.~\ref{fig:full_pareto} of the main text.

\subsubsection{Case 2}
We now consider the case where the term in the square parenthesis of Eq.~(\ref{eq:app_h1}) is zero. This happens in two cases, i.e. when
\begin{equation}
\begin{aligned}
    \delta x_\text{C}^{*2} &=  \frac{a+4b-\sqrt{a^2-8bc}}{2a+4b+2c}dT, \\
    \delta x_\text{C}^{*3} &= \frac{a+4b+\sqrt{a^2-8bc}}{2a+4b+2c}dT.
\end{aligned}
\end{equation}
These solutions hold when $a^2-8bc\geq 0$. In the opposite case, there is no solution. Both solutions can be shown to always satisfy the heat engine condition in Eq.~(\ref{eq:dxc_condition}), and in this point we have
\begin{equation}
    \ev*{G}^{*2,3} = 0.
\end{equation}
However, this zero value of the figure of merit occurs delivering a positive power that is compensated by the negative sign in front of the finite fluctuations and of the finite entropy production. It can be shown that such solutions either lie on the boundary Eq.~(\ref{eq:pareto_analytic}), or inside. We further notice that, looking at the figure of merit, both these solutions are suboptimal with respect to case $1$, except for being equivalent in the special case when $a^2=8bc$.

\subsubsection{Fast driving Pareto front}
As we have seen, Otto cycles in the fast driving, expanded to leading order in the temperature difference, exactly satisfy the steady-state thermodynamic uncertainty relation $\xi=1$. Furthermore, the trade-off between power and power fluctuations is given by Eq.~(\ref{eq:pareto_analytic}). Therefore, the entire Pareto front, plotted in Fig.~\ref{fig:full_pareto}a of the main text, is simply obtained by imposing the thermodynamic uncertainty relation, together with the boundary of Eq.~(\ref{eq:pareto_analytic}).

However, it could in principle be possible that points inside the boundary of Eq.~(\ref{eq:pareto_analytic}) may not be reached, i.e. there may not be any Otto cycle in the fast driving regime reaching some of these points. Using the results of the previous subsections, it can be verified that this is not the case, i.e. that there is an Otto cycle in the fast driving regime corresponding to all points shown in Fig.~\ref{fig:full_pareto}a of the main text.

\subsubsection{Positive figure of merit boundary for Otto cycles in the fast-driving regime}
In this section we derive the expression for the black boundary shown in Fig.~\ref{fig:fig_merit}b of the main text.
As we have seen in the previous two subsections, the condition
\begin{equation}
    a^2=8bc,
    \label{eq:abc_boundary}
\end{equation}
with $b=1-a-c$, determines the boundary between positive and zero value of the optimized figure of merit in the fast-driving regime. However, this condition was found with the normalization of $\ev{P}$, $\ev{\Delta P}$ and $\ev{\Sigma}$ defined as in Eq.~(\ref{eq:g_figmerit_def}), while the boundary shown in Fig.~\ref{fig:fig_merit} of the main text is relative to the normalization chosen in Eq.~(\ref{eq:F_def}) of the main text.
To account for this, we need to ``change the coordinate system'' from $(a_0,b_0,c_0)$ to $(a_1,b_1,c_1)$. Let us define
\begin{equation}
	\ev*{G_0} \equiv a_0 \ev*{P} -b_0 \ev*{\Delta P} - c_0 \ev*{\Sigma},
\end{equation}
and let us consider a given cycle that maximize $G_0$ at given $(a_0,b_0,c_0)$. We want to see how the coefficients $(a_1,b_1,c_1)$ must be chosen in order to find the \textit{same} cycle when maximizing
\begin{equation}
	\ev*{G_1} = a_1 \left(\frac{\ev*{P}}{\lambda_a}\right) -b_1 \left( \frac{\ev*{\Delta P}}{\lambda_b}\right) - c_1 \left(\frac{\ev*{\Sigma}}{\lambda_c}\right),
\end{equation}
where $\lambda_a$, $\lambda_b$ and $\lambda_c$ are given coefficients.

It can be shown that the following identity holds:
\begin{equation}
	\ev*{G_0} = N \ev*{G1},
\end{equation}
where we choose
\begin{equation}
	(a_1,b_1,c_1) = (\lambda_a a_0,\lambda_b b_0,\lambda_c c_0)/N,
	\label{eq:new_coordinate}
\end{equation}
with
\begin{equation}
	N = \lambda_a a_0 + \lambda_b b_0 +\lambda_c c_0.
\end{equation}
Since $N$ is just a proportionality factor, they will share the same maximums. Therefore Eq.~(\ref{eq:new_coordinate}) defines the ``change of coordinates''. In order to transform the boundary in Eq.~(\ref{eq:abc_boundary}) to the ``new coordinate system'', we need to invert Eq.~(\ref{eq:new_coordinate}). Using that $b=1-a-c$ in both coordinate systems, we find
\begin{equation}
	\begin{pmatrix}
		a_0 \\ c_0
	\end{pmatrix}
= \frac{\lambda_b}{a_1\lambda_c(\lambda_b-\lambda_a) + c_1\lambda_a(\lambda_b-\lambda_c) + \lambda_a\lambda_c}
\begin{pmatrix}
	\lambda_c a_1 \\ \lambda_a c_1
\end{pmatrix}.
\end{equation}
Plugging this into Eq.~(\ref{eq:abc_boundary}) yields
\begin{equation}
    a_1^2 = \frac{\lambda_a^2}{\lambda_b\lambda_c}8c_1(1-a_1 - c_1).
\end{equation}
Using Eq.~(\ref{eq:p_dp_max}), we choose
\begin{align}
	\lambda_a &= \frac{g_\text{max}}{16}dT^2, & \lambda_b &= \frac{g_\text{max}}{8} dT^2, & \lambda_c &=\frac{g_\text{max}}{16}dT^2,
\end{align}
and solve for $a$. Retaining the correct solution, we find
\begin{equation}
    a = 2(c-\sqrt{c}).
\end{equation}
This corresponds to the black curve shown in Fig.~\ref{fig:fig_merit}b of the main text.

\subsection{Slow driving regime}
Here we optimize the trade-off between power, entropy production and power fluctuations for the quantum dot engine in the slow-driving regime. For this regime we split the protocol of the cycle into four steps: (i) we fix the bath at a cold temperature $T_\text{C}$ and we slowly vary the dot energy $\epsilon(t) \equiv u(t)E_0$ for a time $\tau_C$ from $\epsilon(0) = \epsilon_A$ to $\epsilon(\tau_C^-) = \epsilon_B $. (ii) We now change the bath temperature to $T_\text{H}$ ($> T_\text{C}$), all while we are changing the energy to $\epsilon_B T_\text{H}/T_\text{C}$ in such a way that the populations of the dot levels remain constant all along the process (because $[H_{u(t)}, H_{u(t')}] =0 \; \forall \, t,t'$). This property allows us to perform this step arbitrarily fast without affecting the objectives. (iii) while keeping the bath temperature at $T_\text{H}$ we slowly vary the dot energy from $\epsilon(\tau_C^+) = \epsilon_B T_\text{H}/T_\text{C}$ to $\epsilon(\tau^-) = \epsilon_A T_\text{H}/T_\text{C}$ in a time $\tau_H = \tau-\tau_C$. (iv) A second quench is performed (in the same manner as step (ii)) to bring the temperature to $T_\text{C}$ and the dot energy to $\epsilon_A$, which closes the cycle.

During each isotherm we can divide the heat exchange in the following manner:
\begin{equation}
    Q_x = T_x \Delta S_x - W_x^{\textrm{diss}}, \quad x=(H,C)
\end{equation}
where $T_x \Delta S_x$ is the reversible contribution and corresponds to the quasistatic limit $\tau_x \rightarrow \infty$. The reversible term is given by $\Delta S \equiv \Delta S_H = -\Delta S_C = S(\pi_0)-S(\pi_{\tau_C})$ with $S(\pi)$ the Von Neumann entropy. Crucially, we will from now on assume the \emph{slow driving} regime (aka. \emph{low dissipation}). That is, we assume that the relaxation time scale $\gamma^{-1}$ of the system is much smaller than the time it takes to complete the protocol. This allows us to expand the relevant quantities in orders of $1/(\gamma\tau_x)$ and keep only the leading terms. In this regime the state can be written as
\begin{equation}
    \rho_t = \pi_t + \frac{1}{\tau\gamma}\rho^{(1)}_t + \mathcal{O}(\frac{1}{\gamma^2\tau^2}).
\end{equation}
It is important to note that when the quenches of steps (ii) and (iv) are performed the adiabatic term $\pi_t$ only contributes reversible work, but the corrective term $\frac{1}{\tau\gamma}\rho^{(1)}_t$ will cause dissipated work of order $\frac{1}{\tau\gamma}$ . For simplicity we want to neglect those terms, in order to do so we can add a waiting time $\tau_w = \tau \Delta s$ at the end of each isotherm. By using the same Lindbladian model as in the fast driving regime \eqref{eq:master_eq_sm}, it is clear that during this waiting time the corrective term is exponentially suppressed. Therefore the total protocol time is now $\tau = \tau_C + \tau_H + 2\tau_w = \frac{\tau_C+\tau_H}{1-2\Delta s}$.

As a result of the slow driving expansion, for each isotherm the irreversible terms can be written as \cite{salamon1983,abiuso2020_prl,scandi2019,schlogl1985,nulton1985,crooks2007,sivak2012,bonanca2014,cavina2017_prl,deffner2020}: 
\begin{equation}
    W_x^{\textrm{diss}} = \frac{T_x}{\tau_x}\int ds~  m_{\epsilon\epsilon}\dot\epsilon(s)^2 \equiv T_x \frac{\sigma_x}{\tau_x} + \mathcal{O}(\frac{1}{\gamma^2\tau_x^2}),
\end{equation}
where $m_{\epsilon\epsilon} = \gamma^{-1}\partial^2\ln\Tr[e^{-\beta H_u}]/\partial\epsilon^2$ is the thermodynamic metric. Furthermore, since we only change the temperature of the bath without affecting its spectral density we have that the optimal protocol verifies $\sigma_H = \sigma_C$ \cite{abiuso2020_prl,cavina2017_prl}. We will therefore drop this index going forward.
In a cycle the work extracted is $W = Q_C + Q_H$, therefore the power of the engine reads
\begin{equation}
    \langle P\rangle = \frac{1}{\tau}\Delta T \Delta S - \frac{\sigma}{\tau}\left(\frac{T_\text{H}}{\tau_H} + \frac{T_\text{C}}{\tau_C}\right),
\end{equation}
with $\Delta = T_\text{H} - T_\text{C}$. The slow driving regime allows us to use fluctuation dissipation relations \cite{weber1956,kubo1966,hermans1991,jarzynski1997,hendrix2001,miller2019} to compute the fluctuations of the work: in particular we have $\frac{1}{2}\textrm{Var}(W_x) = T_x W_x^{\textrm{diss}}$. Instead over the quenches we get $\textrm{Var}(W_y) = \Delta T^2 C_y$, where $C_y = -\beta^2 \partial^2 \ln\Tr[e^{-\beta H_{u_y}}]/\partial \beta^2$ is the heat capacity, for $y=A,B$. Therefore the power fluctuations of the full cycle are
\begin{equation}
    \langle\Delta P\rangle = \frac{\textrm{Var}(W)}{\tau} = \frac{1}{\tau}\Delta T^2 \left(C_A + C_B\right) + \frac{2\sigma}{\tau}\left(\frac{T_\text{H}^2}{\tau_H} + \frac{T_\text{C}^2}{\tau_C}\right).
\end{equation}
The entropy production rate is, by definition,
\begin{equation}
    \langle\Sigma\rangle = \frac{\sigma}{\tau}\left(\frac{1}{\tau_H} + \frac{1}{\tau_C}\right).
\end{equation}
In this appendix we define a figure of merit where the normalization has been absorbed in the weights
\begin{equation}
    G = a\langle P\rangle - \frac{b}{2}\langle\Delta P\rangle - c\langle\Sigma\rangle = \frac{1}{\tau}\left[\Delta T A - \sigma\left(\frac{\delta_H^2}{\tau_H} + \frac{\delta_C^2}{\tau_C}\right)\right]
    \label{eq:fom_SD}
\end{equation}
with $A = a\Delta S - \frac{1}{2}b\Delta T (C_A + C_B)$ and $\delta_x^2 = a T_x + b T_x^2 + c$. 
\subsubsection{Optimization of the cycle}
From \eqref{eq:fom_SD} we can see that, for given bath temperatures, $A$ depends only on the boundary values $\epsilon_A$ and $\epsilon_B$ and $\sigma$ only depends on the protocol's geometric shape between those boundaries. Therefore all the dependence on protocol time has been rendered explicit and we can optimize over it by setting the derivatives with respect to $\tau_C$ and $\tau_C$ to zero: we find
\begin{align}
\tau_H &= 2\left(\delta_H + \delta_C\right)\frac{\delta_H\sigma}{\Delta T A}, \\
\tau_C &= 2 \left(\delta_H + \delta_C\right)\frac{\delta_C\sigma}{\Delta T A}, \\
\tau &= 2\frac{\left(\delta_H + \delta_C\right)^2}{1-2\Delta s}\frac{\sigma}{\Delta T A}.\label{eq:SD_time_tot}
\end{align}
It is important to note that, for the time optimization to be valid, we require that $A$ is strictly positive. If $A$ is negative we can see that $G$ is also negative. But a cycle in which we do nothing results in $G = 0$ which is "better" than a cycle where $G$ is negative, therefore we can always assume $A$ to be positive.
We must now check the positivity of $A$. It is quite easy to see that if $a=0$ then $A$ is negative and therefore the optimal cycle is to do nothing. But if $a>0$ we then have that $A$ is strictly larger than zero if and only if
\begin{equation}\label{eq:cond_A_pos}
    \frac{\Delta S}{C_A + C_B} >  \frac{b}{2a}\Delta T.
\end{equation}
It is clear that this equation can always be satisfied by choosing endpoints of the protocol ($\epsilon_A$ and $\epsilon_B$) that have vanishing heat capacity.\\
When we insert the optimal times into the figure of merit we find
\begin{equation}
    G = \frac{A^2\Delta T^2}{4\sigma}\frac{1-2\Delta s}{(\delta_H + \delta_C)^2}.
\end{equation}
During the waiting time at the end of each isotherm, the decay towards the thermal state goes as $e^{-\tau\gamma\Delta s}$, to be able to neglect the correction we have to set $\tau\gamma\Delta s = r$, where $r$ is some arbitrary $\mathcal{O}(1)$ constant that is sufficiently large.
Therefore we find $\Delta s =  \frac{1}{2} \frac{r A}{r A + (\delta_H + \delta_C)^2\sigma}$, which leads to a new form of $G$:
\begin{equation}
    G = \frac{1}{4}\frac{A^2\Delta T^2}{r A + (\delta_H + \delta_C)^2\sigma}.
\end{equation}
At this point, the only term that depends on the protocol shape (given its boundaries) is $\sigma$. As one would expect, $G$ is maximized when $\sigma$ is minimized. It has been shown \cite{abiuso2020_entropy,jencova2004} that $\sigma$ is minimized by the thermodynamic length $\sigma_{min} = L^2 = 4 \arccos\!\left(\sqrt{p_A p_B} + \sqrt{(1-p_A)(1-p_B)}\right)^2$, where $p_{A,B} = (1+e^{\beta_C \epsilon_{A,B}})^{-1}$ is the occupation probability of the excited state at the endpoints of the isotherms.

\begin{figure}[H]
	\centering
	\begin{minipage}{0.49\textwidth}
        \centering
        \includegraphics[width=0.98\textwidth]{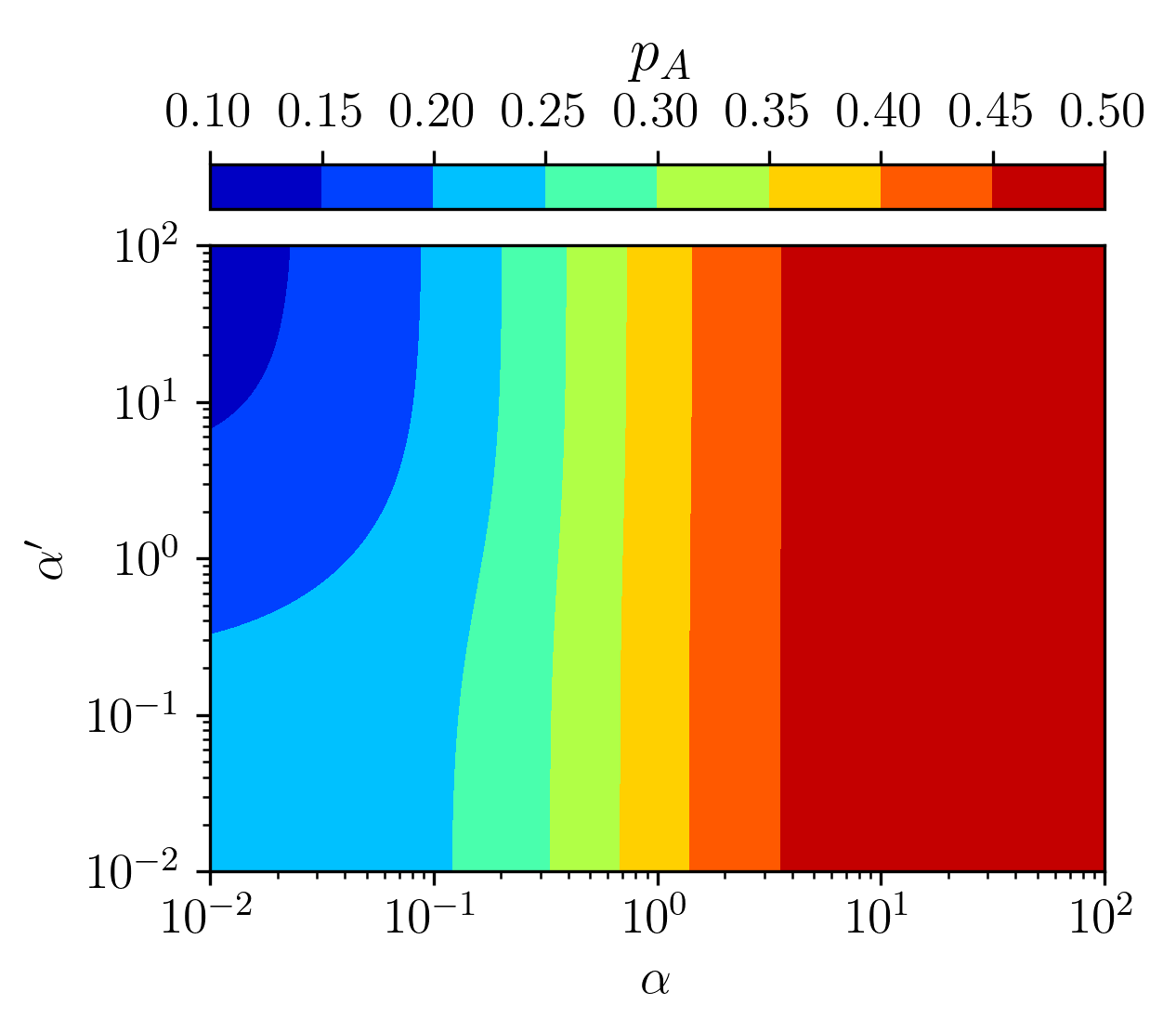}
    \end{minipage}\hfill
    \begin{minipage}{0.49\textwidth}
        \centering
        \includegraphics[width=0.98\textwidth]{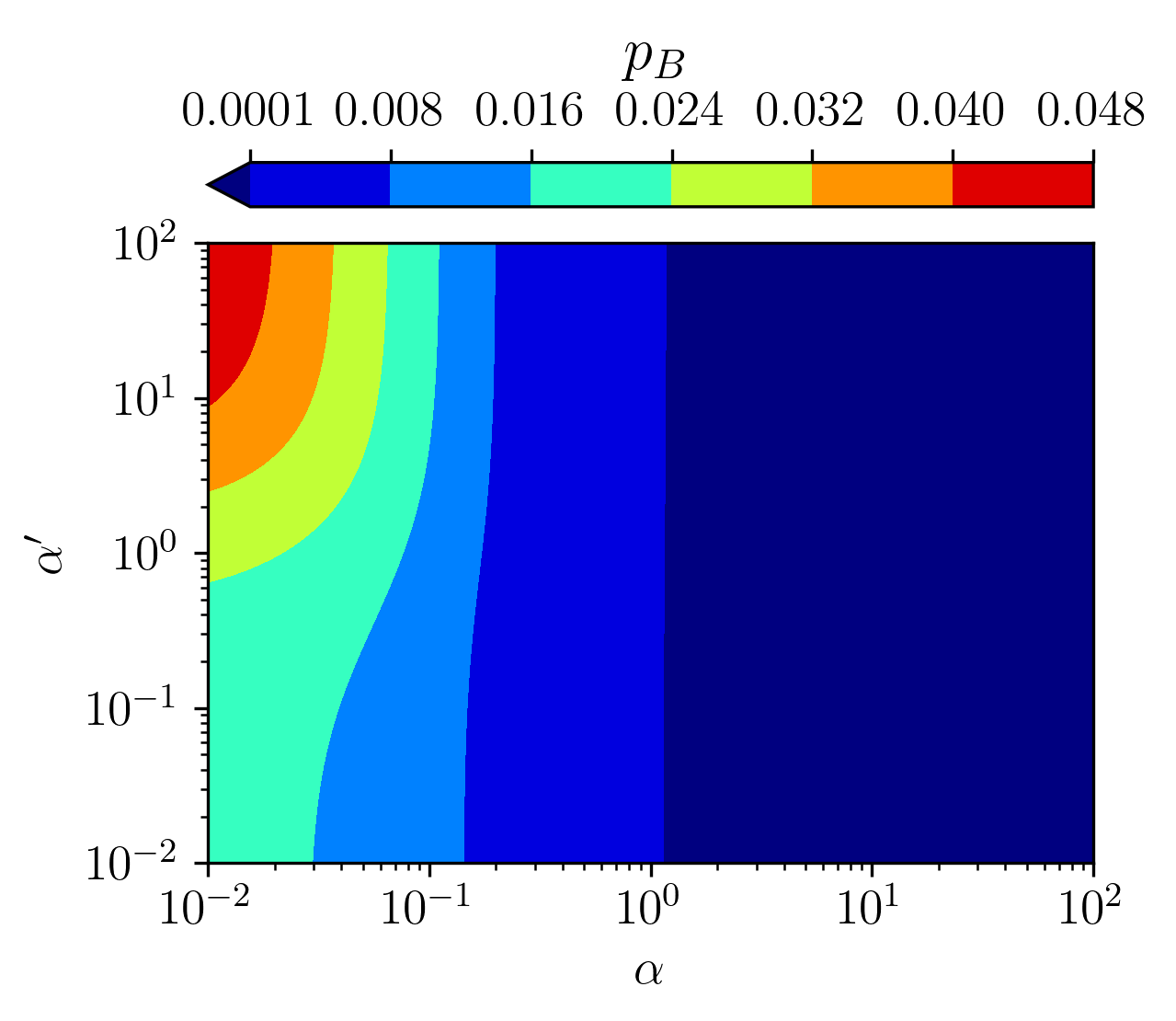}
    \end{minipage}
	\caption{Probabilities of occupation of the excited state at the endpoints of the isotherms ($A$ and $B$) that maximize $G$ in the slow-driving regime. These are displayed as a function of $\alpha = b\Delta T/a$ and $\alpha' = c/a$ (parameters: $\beta_C = 2$, $\beta_H = 1$ and $r=2$). These values are used to produce the Pareto front in panel b of Fig. \ref{fig:full_pareto} of the main text.}
	\label{fig:SD_optimization}
\end{figure}

Now we only have to optimize over the endpoints of the protocol the figure of merit $G = \frac{1}{4}\frac{A^2\Delta T^2}{r A + (\delta_H + \delta_C)^2 L^2}$. By taking derivatives with respect to $\epsilon_A$ and $\epsilon_B$ we find two transcendental equations that cannot be solved analytically. Equivalently, we can optimize with respect to $p_A$ and $p_B$, though the same problem persists. But these variables are more practical for numerical optimization as they are defined on the finite range $[0,0.5]$.
Therefore we do a numerical maximization of $G$ for this last step, from the form of $G$ it can be seen that the couple $(p_A,p_B)$ that maximizes $G$ depends only on two parameters (assuming the temperatures are given): $\alpha := b \Delta T/a$ and $\alpha' := c/a$. We can note that a choice of these parameters defines which trade-off of the objectives we want. Furthermore we have to satisfy the constraint $\frac{\Delta S}{C_A+C_B} > \alpha/2$ (see Eq. \eqref{eq:cond_A_pos}). From a numerical point of view, in order to plot the Pareto front, we are interested in a range of values of $\alpha$ and $\alpha'$ that range from $10^{-2}$ to $10^{2}$. The optimization was done using a \emph{Mathematica} script and is shown in Fig. \ref{fig:SD_optimization}, then we can use this to compute the power, efficiency, and power fluctuations for different trade-off ratios $\alpha$ and $\alpha'$ with which we obtain the Pareto front shown in panel b of Fig. \ref{fig:full_pareto} of the main text.

\subsubsection{Violation of thermodynamic uncertainty relation.}
In the low power limit it is known that PDHE can violate thermodynamic uncertainty relations \cite{barato2016,proesmans2017,holubec2018,cangemi2020,menczel2021,lu2022}. We reach this limit by not optimizing for power but instead prioritizing the other two objectives.
We take the limit of low power by taking $a\ll b,c$; this implies that $\alpha, \alpha' \gg 1$, by Eq. \eqref{eq:cond_A_pos} this implies that we need vanishing heat capacity at $A$ and $B$ (because $\Delta S$ is upper-bounded by $\ln 2$). This is seen in Fig. \ref{fig:SD_optimization}: $p_A$ approaches the largest values it can while $p_B$ approaches the smallest values it can, which both minimize heat capacity. Furthermore, this also implies that the thermodynamic distance is as large as it can be (since it is a measure of the distance between the states at $A$ and $B$), but crucially it remains finite. By applying this to Eq. \eqref{eq:SD_time_tot} it is clear that the magnitude of the protocol time is determined by the ratio $\frac{L^2}{A} =  \frac{1}{a}\frac{L^2}{\Delta S - \alpha(C_A + C_B)/2}$, it is thus clear that the protocol time diverges in this limit. 

The fact that the protocol time diverges in this limit also explains the fact that for large $\alpha$ we see a loss of the dependence on $\alpha'$ in Fig. \ref{fig:SD_optimization}: in this limit we don't need to fix $\tau\gamma\Delta s$ to a constant, we only need to fix $\Delta s$ to an arbitrarily low constant since $\tau$ goes to infinity. This allows us to keep a simpler form of the figure of merit:  $G = \frac{A^2\Delta T^2}{4 L^2}\frac{1-2\Delta s}{(\delta_H + \delta_C)^2}$, in which we can note that the optimization over the protocol endpoints only has to be done over the ratio $A^2/L^2$; that does not depend on $\alpha'$. To maximize this ratio we can take the derivatives with respect to $p_A$ and $p_B$. As before we end up with transcendental equations: 
\begin{align}
\frac{\alpha^{-1} \Delta S - (C_A+C_B)/2}{\Delta T L} &= \beta_C\epsilon_A\sqrt{p_A(1-p_A)} \left[\alpha^{-1} + 1-\frac{1}{2}\beta_C\epsilon_A(1-2 p_A) \right]  , \\
\frac{\alpha^{-1} \Delta S - (C_A+C_B)/2}{\Delta T L} &= \beta_C\epsilon_B\sqrt{p_B(1-p_B)} \left[\alpha^{-1} - 1+\frac{1}{2}\beta_C\epsilon_B(1-2 p_B) \right] .
\end{align}
For $\alpha^{-1}=0$ we can see that $p_A = \frac{1}{2}$ and $p_B = 0$ solves the equations, which is coherent with the numerical results of the previous section. We can now expand around $\alpha^{-1}=0$ by setting $p_A = \frac{1}{2} + p_A^{(1)}\alpha^{-1} + \mathcal{O}(\alpha^{-2})$ and $p_B = 0 + p_B^{(1)}\alpha^{-1} + \mathcal{O}(\alpha^{-2})$. We therefore find
\begin{align}
p_A &= \frac{1}{2} - \frac{\ln 2}{\pi}\alpha^{-1} + \mathcal{O}(\alpha^{-2}) , \\
p_B &= \mathcal{O}(\alpha^{-2}) .
\end{align}
With this result we can compute the values of the objectives analytically to leading order in the low power limit:
\begin{align}
    \langle P\rangle &= 2\alpha^{-1} \left(\frac{\ln 2}{\pi}\right)^2 \frac{\Delta T^2}{\left(\sqrt{\frac{c}{b}+T_\text{H}^2} + \sqrt{\frac{c}{b}+T_\text{C}^2}\right)^2} + \mathcal{O}(\alpha^{-2}), \\
    \langle\Delta P\rangle &= 2\alpha^{-2} \left(\frac{\ln 2}{\pi}\right)^2 \frac{\Delta T^2}{\left(\sqrt{\frac{c}{b}+T_\text{H}^2} + \sqrt{\frac{c}{b}+T_\text{C}^2}\right)^3}\left(\frac{T_\text{H}^2}{\sqrt{\frac{c}{b}+T_\text{H}^2}} +\frac{T_\text{C}^2}{\sqrt{\frac{c}{b}+T_\text{C}^2}}\right) + \mathcal{O}(\alpha^{-3}), \\
    \langle\Sigma\rangle &= \alpha^{-2}\left(\frac{\ln 2}{\pi}\right)^2\frac{1}{\sqrt{\frac{c}{b}+T_\text{H}^2}\sqrt{\frac{c}{b}+T_\text{C}^2}}~ \frac{\Delta T^2}{\left(\sqrt{\frac{c}{b}+T_\text{H}^2} + \sqrt{\frac{c}{b}+T_\text{C}^2}\right)^2} + \mathcal{O}(\alpha^{-3}).
\end{align}
In particular we notice that $P \propto \alpha^{-1}$, $\Delta P \propto \alpha^{-2}$, $\langle\Sigma\rangle \propto \alpha^{-2}$. Which results in $\xi \propto \alpha^2$, therefore we can violate the thermodynamic uncertainty relations arbitrarily. Therefore we can write, in this limit, $\xi$ in terms of $\langle P\rangle$, $\langle\Delta P\rangle$ and $\langle\Sigma\rangle$
\begin{align}
    \xi &=  \frac{16 \left(\frac{\ln 2}{\pi}\right)^4 P^{-2} \Delta T^4}{\left(\sqrt{\frac{c}{b}+T_\text{H}^2} + \sqrt{\frac{c}{b}+T_{c\vphantom{h}}^2}\right)^3} \frac{\left(\frac{c}{b}+T_\text{H}^2\right)\left(\frac{c}{b}+T_{c\vphantom{h}}^2\right)}{T_\text{C}^2\sqrt{\frac{c}{b}+T_\text{H}^2} + T_\text{H}^2\sqrt{\frac{c}{b}+T_{c\vphantom{h}}^2}},\\
    \xi &= \frac{4 \left(\frac{\ln 2}{\pi}\right)^2 \langle\dot\Sigma\rangle^{-1} \Delta T^2}{\sqrt{\frac{c}{b}+T_\text{H}^2} + \sqrt{\frac{c}{b}+T_{c\vphantom{h}}^2}}~ \frac{\sqrt{\frac{c}{b}+T_\text{H}^2}\sqrt{\frac{c}{b}+T_{c\vphantom{h}}^2}}{T_\text{C}^2\sqrt{\frac{c}{b}+T_\text{H}^2} + T_\text{H}^2\sqrt{\frac{c}{b}+T_{c\vphantom{h}}^2}}, \\
    \xi &= 8 \left(\frac{\ln 2}{\pi}\right)^2 \Delta P^{-1} \Delta T^2 \frac{\sqrt{\frac{c}{b}+T_\text{H}^2}\sqrt{\frac{c}{b}+T_{c\vphantom{h}}^2}}{\left(\sqrt{\frac{c}{b}+T_\text{H}^2} + \sqrt{\frac{c}{b}+T_{c\vphantom{h}}^2}\right)^2}.
\end{align}
By maximising over the choice of $b$ and $c$ for a given $\alpha$ we can then obtain the black lines in Fig. \ref{fig:tur} of the main text.

\subsection{Numerical Optimization using RL}
Here we provide the training details and hyperparameters used to produce all RL results presented in the main text in Figs.~\ref{fig:fig_merit}, \ref{fig:full_pareto}, and \ref{fig:tur} of the main text. The method is described in Sec.~\ref{sec:rl}. A separate training was performed for all values of the weights $(a,c)$ reported in Fig.~\ref{fig:fig_merit} of the main text. Since the cycles vary dramatically based on the choice of $(a,c)$, we mainly employed two sets of hyperparameters, denoted with v1 and v2, shown in Table~\ref{tab:hyper}. The hyperparameter names in Table~\ref{tab:hyper} not defined in this paper are defined as in Ref.~\cite{erdman2022}, except for $n_\text{train-steps}$ that represents the number of steps that were performed during training. For values of $a$ in $[0.4,1]$, we used v1 and performed a single training. The only two exceptions are $(a,c) = (1,0)$ and $(0.55,0)$, where we obtained better cycles training a second time respectively with $n_\text{train-steps}=340k$ and $280k$. 
Instead, training for values of $a$ in $[0.2,0.35]$ was less stable, since the low power regime produces much longer cycles that require more ``long-term planning''. Furthermore, the maximum value of the figure of merit becomes closer to $0$ as $a$ decreases. Since also ``doing nothing'' (i.e. setting $\vec{u}(t)$ and $\beta(t)$ to a time-independent constant) produces a null figure of merit, it becomes harder and harder for the RL method to distinguish optimal cycles from these trivial solutions. 
To overcome these difficulties, we mainly used v2, and we repeated the training up to $3$ times for some values of $(a,c)$ - choosing then the cycle with the largest figure of merit. In total, we repeated $54$ trainings, some of which trained with v1. At last, in some cases we trained for more steps, up to $n_\text{train-steps}=471k$ steps, and we set $\Delta t=0.5$ in some other trains, including the cases with $c=0$ or $c=1-a$.
\begin{table}[h]
\centering
\begin{tabular}{lll}
\toprule    
 Hyperparameter name \qquad ~  &  v1 \qquad ~ &   v2
 \\\colrule 
 Hidden layers & 2 & " \\
 Hidden layers units & 256 & "  \\
 Initial random steps & 6k & " \\
 First update at step & 1000 & " \\
 Batch size & 256 & " \\
 Learning rate & 0.001 & " \\
 $\bar{H}_\text{start}$ & 0.4 & 0.28 \\
 $\bar{H}_\text{end}$ & -7 & " \\
 $\bar{H}_\text{decay}$ & 108k & 162k \\
 $n_\text{updates}$ & 50 & " \\
 $\rho_\text{polyak}$ & 0.995 & " \\
 $\mathcal{B}_\text{size}$ & 200k & " \\
 $\Delta t$ & 0.5 &  2 \\
 Discount factor $\gamma$ & 0.9997 & 0.9998 \\
 $n_\text{train-steps}$ &  240k & 360k 
  \\\botrule    
\end{tabular}
\caption{Hyperparameters used to produce the results shown in the main text. The columns v1 and v2 correspond to two different sets of hyperparameters used in different regimes. The values not specified in v2 are the same as in v1.}
\label{tab:hyper}
\end{table}

\subsection{Equivalence between SSHE and Fast-Otto PDHE for a two-level system}
In this section we show that there is mapping between power, fluctuations, and entropy production produced by a two-level system in a steady-state heat engine (SSHE) and the same system periodically driven (PDHE) in a fast-Otto cycle.
This affinity can be explained, intuitively speaking, as in both cases the state of the system relaxes to a fixed point (cf.~\cite{cavina2021}), and the work exchanges are defined by a single variation of energy level (chemical potential for the SSHE, energy gap quench for the PDHE, see details below).

This mapping explains why, among other things, the Fast-Otto cycle satisfies the SSHE thermodynamic uncertainty relations, as shown in the main text.

Specifically, consider a two-level system connected to two thermal baths, modeled via a simple rate master equation, that is
\begin{align}
    \dot{\vec{p}}_t=\gamma_i(\vec{\pi}_i-\vec{p}_t)
\end{align}
where $\gamma_i$ is the rate of bath $i$, $\vec{p}$ the populations vector, and $\vec{\pi}_i:=(f_i,1-f_i)$, with $f_i=\frac{1}{1+e^{-\beta_i\epsilon_i}}$. Notice that due to normalization $\vec{p}=(p,1-p)$ and the dynamical equation can be equivalently written as $\dot{p}_t=\gamma_i (f_i-p_t)$. Both in the steady state case and the fast driving cycle, the state tends to a fixed value
\begin{align}
    p_t\rightarrow p\;.
\end{align}

It is possible to compute the steady state $p$, and the resulting power, efficiency and fluctuations for such system when driven on a Fast-Otto cycle, made by two quenches rapidly alternating. The analytic expressions can be obtained as from Appendix~\ref{app:Fast_Driving} (cf. Eq.~\eqref{eq:qd_otto}, but we also allow different rates for the two baths now).

The resulting expressions are given in Table~\ref{tab:mapping_qubit} (right column), in terms of the rates $\gamma_i$, the chosen gaps $\epsilon_i$, the fraction of time $\theta_i$ spent on each bath $i$ ($\theta_1+\theta_2=1$).
It turns out that such expressions become formally equivalent to those of a SSHE based on a two-level system in a chemical potential gradient between the two baths $\mu_i$ (the gap $\epsilon$ of the qubit is in this case fixed), for which the standard expressions in the left column of Table~\ref{tab:mapping_qubit} can be obtained from standard references. In particular, for $\Delta P$ it is possible to consider Eq.~(14) of~\cite{Liu2019} and substitute Eq.~(24) of~\cite{Agarwalla2018a} for the values of the $\mathcal{T}$-coefficients in the weak coupling limit.

By looking at Table~\ref{tab:mapping_qubit}, it is possible to see how each quantity is formally equivalent between the two columns, via the following mapping
\begin{align}
    \gamma_i &\rightarrow \theta_i\gamma_i\;, \\
    \epsilon-\mu_i &\rightarrow \epsilon_i\;.
\end{align}
Finally, let us notice that the expression for the fluctuations $\Delta P$ can equivalently be expressed in other forms, such as (we express it for the SSHE case, although it can be expressed similarly for the Fast-Otto case, via the above mapping)
\begin{align}
    \Delta P = \frac{2\gamma_1\gamma_2}{\gamma_1+\gamma_2}(\mu_1-\mu_2)^2\left(f_+(1-f_+)+(\bar{f}-f_+)^2\right)\;,
\end{align}
where $f_+:=(f_1+f_2)/2$. Such expression shows immediately the positiveness of the fluctuations. Another equivalent expression, which is typically obtained in SSHE calculations is
\begin{align}
    \Delta P= \frac{\gamma_1\gamma_2}{\gamma_1+\gamma_2}(\mu_1-\mu_2)^2\left(f_1(1-f_2)+f_2(1-f_1)\right) -\frac{2\gamma_1^2\gamma_2^2}{(\gamma_1+\gamma_2)^3}(\mu_1-\mu_2)^2\left(f_1-f_2\right)^2\;.
\end{align}

\begin{table}
\label{tab:mapping_qubit}
\begin{tabular}{c|c|c}
    & SSHE & Fast-Otto PDHE   \\ \hline
    $f_i$ & $(1+e^{-\beta_i(\epsilon-\mu_i)})^{-1}$ &  $(1+e^{-\beta_i\epsilon_i})^{-1}$ \\
   $p$ & $\bar{f}:=\dfrac{\gamma_1 f_1+\gamma_2 f_2}{\gamma_1+\gamma_2}$   & $\bar{f}:=\dfrac{\theta_1\gamma_1 f_1+\theta_2\gamma_2 f_2}{\theta_1\gamma_1+\theta_1\gamma_2}$  \\
    $P$ & $\dfrac{\gamma_1\gamma_2}{\gamma_1+\gamma_2}(f_1-f_2)(\mu_1-\mu_2) $ &  $\dfrac{\theta_1\theta_2\gamma_1\gamma_2}{\theta_1\gamma_1+\theta_2\gamma_2}(f_1-f_2)(\epsilon_2-\epsilon_1) $\\
    $\eta$ & $1-\dfrac{\epsilon-\mu_1}{\epsilon-\mu_2}$ & $1-\dfrac{\epsilon_1}{\epsilon_2}$\\
   $\Delta P$ &  $2\frac{\gamma_1\gamma_2}{\gamma_1+\gamma_2}
     (\epsilon_1-\epsilon_2)^2
    \left[ \bar{f} (1-\bar{f}) + \left(\frac{f_1+f_2}{2} -\bar{f}\right)(1-2\bar{f}) \right]$ & 
   $2\frac{\gamma_1\theta_1\gamma_2\theta_2}{\gamma_1\theta_1+\gamma_2\theta_2}
     (\epsilon_1-\epsilon_2)^2
    \left[ \bar{f} (1-\bar{f}) + \left(\frac{f_1+f_2}{2} -\bar{f}\right)(1-2\bar{f}) \right]$
\end{tabular}
\caption{Mapping between a steady-state engine and a fast-Otto cycle for the case of a two-level system.}
\end{table}

\end{document}